\documentstyle[aps,prd,epsfig,preprint]{revtex}
\def\fun#1#2{\lower3.6pt
\vbox{\baselineskip0pt\lineskip.9pt
\ialign{$\mathsurround=0pt#1\hfill##\hfil$
\crcr#2\crcr\sim\crcr}}}
\def\Figa{1}
\def\Figb{2}
\def\Figc{3}
\def\Figd{4}
\def\Fige{5}
\def\Figf{6}
\def\Figg{7}
\def\Figh{8}
\def\Figi{9}
\def\Figj{10}
\begin{document}
\vspace{0.5in}
\title{\vskip-2.5truecm{\hfill \baselineskip 14pt
{\hfill {{\small \hfill UT-STPD-1/02}}}\\
{{\small \hfill FISIST/05-2002/CFIF}}
\vskip .1truecm}
\vspace{1.0cm}
\vskip 0.1truecm{\bf Yukawa Quasi-Unification}}
\vspace{1.0cm}
\author{{M.E. G\'{o}mez}$^{(1)}$\thanks{mgomez@cfif.ist.utl.pt},
{G. Lazarides}$^{(2)}$\thanks{lazaride@eng.auth.gr}
{and C. Pallis}$^{(3)}$\thanks{pallis@he.sissa.it}}
\vspace{1.0cm}
\address{$^{(1)}${\it Centro de F\'{\i}sica das
Interac\c{c}\~{o}es Fundamentais (CFIF),
Departamento de F\'{\i}sica, \\ Instituto Superior T\'{e}cnico,
Av. Rovisco Pais, 1049-001 Lisboa, Portugal.}}
\address{$^{(2)}${\it Physics Division, School of Technology,
Aristotle University of Thessaloniki,\\
54124 Thessaloniki, Greece.}}
\address{$^{(3)}${\it Scuola Internazionale Superiore Di Studi
Avanzati (SISSA), \\
Via Beirut 2-4, 34013 Trieste, Italy.}} \maketitle

\vspace{.5cm}

\begin{abstract}
\baselineskip 12pt

\par
We construct concrete supersymmetric grand unified
theories based on the Pati-Salam gauge group
$SU(4)_c\times SU(2)_L\times SU(2)_R$ which naturally
lead to a moderate violation of `asymptotic' Yukawa
unification and thus can allow an acceptable
$b$-quark mass even with universal boundary
conditions. We consider the constrained minimal
supersymmetric standard model which emerges from
one of these theories with a deviation from Yukawa
unification which is adequate for $\mu>0$. We show
that this model possesses a wide and natural range
of parameters which is consistent with the data on
$b\rightarrow s\gamma$, the muon anomalous magnetic
moment, the cold dark matter abundance in the
universe, and the Higgs boson masses. The lightest
supersymmetric particle can be as light as about
$107~{\rm GeV}$.
\end{abstract}

\thispagestyle{empty}
\newpage
\pagestyle{plain}
\setcounter{page}{1}
\baselineskip 20pt

\section{Introduction}
\label{intro}

\par
The most restrictive version of the minimal supersymmetric
standard model (MSSM) with gauge coupling unification is
based on radiative electroweak breaking with universal
boundary conditions from gravity-mediated soft
supersymmetry (SUSY) breaking and is known as constrained
MSSM (CMSSM). It is desirable to further restrict this
model by assuming that the $t$-quark, $b$-quark and
$\tau$-lepton Yukawa couplings unify `asymptotically',
i.e., at the SUSY grand unified theory (GUT) scale
$M_{GUT}\approx 2\times 10^{16}~{\rm GeV}$. This assumption
(Yukawa unification) naturally restricts \cite{als} the
$t$-quark mass to large values compatible with the data.
Also, the emerging model is highly predictive.

\par
Yukawa unification can be achieved by embedding the MSSM
in a SUSY GUT with a gauge group containing $SU(4)_c$ and
$SU(2)_R$. Indeed, assuming that the electroweak Higgs
superfields $h^{ew}_1$, $h^{ew}_2$ and the third family
right handed quark superfields $t^c$, $b^c$ form $SU(2)_R$
doublets, we obtain \cite{pana} the `asymptotic' Yukawa
coupling relation $h_t=h_b$ and, hence, large
$\tan\beta\approx m_{t}/m_{b}$. Moreover, if the third
generation quark and lepton $SU(2)_L$ doublets (singlets)
$q_3$ and $l_3$ ($b^c$ and $\tau^c$) form a $SU(4)_c$
4-plet ($\overline{4}$-plet) and the Higgs doublet
$h^{ew}_1$ which couples to them is a $SU(4)_c$ singlet,
we get $h_b=h_{\tau}$ and the `asymptotic' relation
$m_{b}=m_{\tau}$ follows. The simplest GUT gauge group
which contains both $SU(4)_c$ and $SU(2)_R$ is the
Pati-Salam (PS) group $G_{PS}=SU(4)_c\times SU(2)_L\times
SU(2)_R$. Higher groups with the same property are
$SO(10)$ or $E_{6}$.

\par
One problem, which is faced in trying to incorporate Yukawa
unification into the CMSSM, is due to the generation of
sizeable SUSY corrections to $m_b$ \cite{copw,hall} which
have the same sign as the MSSM parameter $\mu$. Consequently,
for $\mu>0$, the tree-level $m_b(M_Z)$, which is predicted
\cite{cd2} (the sign of $\mu$ in this reference is opposite
to the one adopted here) from Yukawa unification already
above the $95\%$ confidence level (c.l.) experimental range
$2.684-3.092~{\rm GeV}$, receives large positive corrections.
This range is derived from the $95\%$ c.l. range for
$m_b(m_b)$ in the $\overline{MS}$ renormalization scheme
which is found \cite{mb} (see also Ref.\cite{cmssm}) to be
$3.95-4.55~{\rm GeV}$. This is evolved \cite{santamaria} up
to $M_Z$ via the three-loop $\overline{MS}$ renormalization
group (RG) equations \cite{rge} with $\alpha_s(M_Z)=0.1185$
and then converted to the $\overline{DR}$ scheme using the
appropriate one-loop factor \cite{dr}. For $\mu<0$, the
tree-level $m_b(M_Z)$ is \cite{cd2} smaller and close to the
upper edge of the above range. However, the radiative
corrections are now negative and drive $m_b(M_Z)$ below this
range. The discrepancy is, though, considerably smaller than
in the $\mu>0$ case.

\par
We see that, for both signs of $\mu$, the hypothesis of
exact Yukawa unification leads to an unacceptable $b$-quark
mass. However, we are not obliged to abandon Yukawa
unification altogether. We can rather modestly correct it by
including an extra $SU(4)_c$ non-singlet Higgs superfield
with Yukawa couplings to the quarks and leptons. The Higgs
$SU(2)_L$ doublets contained in this superfield can naturally
develop \cite{wetterich} subdominant vacuum expectation values
(vevs) and mix with the main electroweak doublets which are
assumed to be $SU(4)_c$ singlets and form a $SU(2)_R$ doublet.
This mixing can, in general, violate $SU(2)_R$. Consequently,
the resulting electroweak Higgs doublets $h_1^{ew}$,
$h_2^{ew}$ do not form a $SU(2)_R$ doublet and also break
$SU(4)_c$. The required deviation from exact Yukawa
unification is expected to be more pronounced in the $\mu>0$
case. Despite this, we choose to study here this case
since, for $\mu<0$, the present experimental data \cite{cleo}
on the inclusive decay $b\rightarrow s\gamma$ restrict
\cite{cd2} the sparticle masses to considerably higher values
and, thus, this case is phenomenologically less interesting.
Moreover, the recent results \cite{muon} on the muon
anomalous magnetic moment also imply heavy sparticles
for $\mu<0$ (see below).

\par
We will construct here a concrete SUSY GUT model which
naturally leads to a modest deviation from Yukawa
unification allowing an acceptable $m_b(M_Z)$ even with
universal boundary conditions. (For models which
violate universality rather than Yukawa unification see
Ref.\cite{nonuniversal}.) We will then show that this
model possesses a wide range of parameters which is
consistent with all the phenomenological and
cosmological constraints. We consider only the $\mu>0$
case, which is experimentally more attractive.

\par
In Sec.\ref{model}, we construct a SUSY GUT model which
is based on the PS gauge group and provides, in a natural
way, a suppressed violation of Yukawa unification.
Variants of this model which can yield bigger deviations
from Yukawa unification are also presented. In
Sec.\ref{mssm}, we concentrate on one of these variants
which can violate Yukawa unification by an amount that
is adequate for $\mu>0$. We then describe the resulting
MSSM under the assumption of universal boundary
conditions and introduce the various phenomenological and
cosmological requirements which restrict its parameter
space. In Sec.\ref{parameters}, we study the range of
parameters which is compatible with all these
requirements in this particular CMSSM. Finally, in
Sec.\ref{conclusions}, we summarize our conclusions.

\section{The SUSY GUT Model}
\label{model}

\par
We take the SUSY GUT model of Ref.\cite{jean} (see
also Ref.\cite{talks}) as our starting point. This is
based on the PS gauge group $G_{PS}$, which is the
simplest gauge group that can lead to Yukawa
unification. The `matter' superfields are
$F_i=(4,2,1)$ and $F^c_i=(\bar{4},1,2)$ ($i=1,2,3$),
while the electroweak Higgs doublets belong to the
superfield $h=(1,2,2)$. So, all the requirements for
exact Yukawa unification are fulfilled. The breaking of
$G_{PS}$ down to the standard model (SM) gauge group
$G_S$ is achieved by the superheavy vevs ($\sim M_{GUT}$)
of the right handed neutrino type components of a
conjugate pair of Higgs superfields $\bar{H}^c=(4,1,2)$,
$H^c=(\bar{4},1,2)$. The model also contains a gauge
singlet $S$ which triggers the breaking of $G_{PS}$,
a $SU(4)_c$ 6-plet $G=(6,1,1)$ which gives
\cite{leontaris} masses to the right handed down quark
type components of $\bar{H}^c$, $H^c$, and a pair of
gauge singlets $\bar{N}$, $N$ for solving \cite{rsym}
the $\mu$ problem of the MSSM via a Peccei-Quinn (PQ)
symmetry. In addition to $G_{PS}$, the model possesses
two global $U(1)$ symmetries, namely a PQ and a R
symmetry, as well as a $Z_2^{mp}$ symmetry (`matter
parity') under which $F$, $F^c$ change sign. For
details on the charge assignments, the full
superpotential and the phenomenological and cosmological
properties of this model, the reader is referred to
Ref.\cite{jean}.

\par
A moderate violation of Yukawa unification can be
accommodated in this model by adding a new Higgs
superfield $h^{\prime}=(15,2,2)$ with Yukawa couplings
$FF^ch^{\prime}$. Actually, this is the only
representation, besides (1,2,2), which possesses such
couplings to the fermions. The existence of these
couplings requires that the quantum numbers of
$h^{\prime}$ coincide with the ones of $h$. So, its PQ
and R charges are $PQ=1$ and $R=0$ respectively. In
order to give superheavy masses to the color
non-singlet components of $h^{\prime}$, we need to
include one more Higgs superfield $\bar{h}^{\prime}=
(15,2,2)$ with the superpotential coupling
$\bar{h}^{\prime}h^{\prime}$, whose coefficient is
of the order of $M_{GUT}$. The field $\bar{h}^{\prime}$
then has $PQ=-1$ and $R=1$. The full superpotential
which is consistent with all the symmetries contains,
in addition to the couplings mentioned above and in
Ref.\cite{jean}, the following extra terms too:
\begin{eqnarray}
FF\bar{H}^c\bar{H}^chh^{\prime},
~FF\bar{H}^c\bar{H}^ch^{\prime}h^{\prime},
~FFH^cH^chh^{\prime},~FFH^cH^ch^{\prime}h^{\prime},
~(\bar{H}^c)^4\bar{h}^{\prime}h,
~(H^c)^4\bar{h}^{\prime}h,
\nonumber \\
~\bar{H}^cH^c\bar{h}^{\prime}h,
~N^2(\bar{H}^c)^4hh^{\prime},~N^2(H^c)^4hh^{\prime},
~N^2\bar{H}^cH^chh^{\prime},~N^2h^{\prime}h^{\prime}.
~~~~~~~~~~~~~
\label{couplings}
\end{eqnarray}
Note that all the superpotential terms can be multiplied
by arbitrary powers of the combinations $(\bar{H}^c)^4$,
$(H^c)^4$, $\bar{H}^cH^c$. The first two couplings in
Eq.(\ref{couplings}) (as well as all the `new' couplings
containing $(\bar{H}^c)^4$, $(H^c)^4$) give rise to
additional baryon and lepton number violation. However,
their contribution to proton decay is subdominant to the
one discussed in Ref.\cite{jean} and proton remains
practically stable.

\par
Let us now discuss how the introduction of
$\bar{h}^{\prime}$, $h^{\prime}$ can lead to a moderate
correction of exact Yukawa unification. The Higgs field
$h^{\prime}$ contains two color singlet $SU(2)_L$
doublets $h_1^{\prime}$, $h_2^{\prime}$ which, after
the breaking of $G_{PS}$ to $G_S$, can mix with the
corresponding doublets $h_1$, $h_2$ in $h$. This is
mainly due to the terms $\bar{h}^{\prime}h^{\prime}$,
$\bar{H}^cH^c\bar{h}^{\prime}h$. The latter, being
non-renormalizable, is suppressed by the string scale
$M_S\approx 5\times 10^{17}~{\rm GeV}$. Actually,
$\bar{H}^cH^c\bar{h}^{\prime}h$ corresponds to
two independent couplings. One of them is between the
$SU(2)_R$ singlets in $\bar{H}^cH^c$ and
$\bar{h}^{\prime}h$, and the other between the $SU(2)_R$
triplets in these combinations. The singlet (triplet)
coupling, after the breaking of $G_{PS}$ to $G_S$,
generates $SU(2)_R$ invariant (violating) bilinear
terms between the doublets in $\bar{h}^{\prime}$ and $h$.
So, we finally obtain two bilinear terms
$\bar{h}_1^{\prime}h_1$ and $\bar{h}_2^{\prime}h_2$ with
different coefficients, which are suppressed by
$M_{GUT}/M_S$. These terms together with the terms
$\bar{h}_1^{\prime}h_1^{\prime}$ and
$\bar{h}_2^{\prime}h_2^{\prime}$ from
$\bar{h}^{\prime}h^{\prime}$, which have equal
coefficients, generate different mixings between $h_1$,
$h_1^{\prime}$ and $h_2$, $h_2^{\prime}$. Consequently,
the resulting electroweak doublets $h_1^{ew}$, $h_2^{ew}$
contain $SU(4)_c$ violating components suppressed by
$M_{GUT}/M_S$ and fail to form a $SU(2)_R$ doublet by an
equally suppressed amount. So, Yukawa unification is
moderately violated. Unfortunately, as it turns out, this
violation is not adequate for correcting the $b$-quark
mass for $\mu>0$.

\par
In order to allow for a more sizable violation of Yukawa
unification, we further extend the model by including a
superfield $\phi=(15,1,3)$ with the coupling
$\phi\bar{h}^{\prime}h$. The field $\phi$ is then neutral
under both the PQ and R symmetries. To give superheavy
masses to the color non-singlets in $\phi$, we introduce
one more superfield $\bar\phi=(15,1,3)$ with the coupling
$\bar\phi\phi$, whose coefficient is of order $M_{GUT}$.
The charges of $\bar\phi$ are $PQ=0$, $R=1$. Additional
superpotential couplings which are allowed by the
symmetries of the model are:
\begin{equation}
\bar\phi(\bar{H}^c)^4,~\bar\phi(H^c)^4,
~\bar\phi\bar{H}^cH^c,~\phi N^2hh^{\prime}.
\label{phicouplings}
\end{equation}
Multiplications of the superpotential terms by $\phi$,
$(\bar{H}^c)^4$, $(H^c)^4$, $\bar{H}^cH^c$ are generally
allowed with the exception of the multiplication of the
$G_{PS}$ singlets $S$, $\bar{N}^2N^2$, $N^2h^2$ by a
single $\phi$, which is a $SU(4)_c$ 15-plet.

\par
The terms $\bar\phi\phi$ and $\bar\phi\bar{H}^cH^c$ imply
that, after the breaking of $G_{PS}$ to $G_S$, the field
$\phi$ acquires a superheavy vev of order $M_{GUT}$. The
coupling $\phi\bar{h}^{\prime}h$ then generates $SU(2)_R$
violating unsuppressed bilinear terms between the doublets
in $\bar{h}^{\prime}$ and $h$. These terms can certainly
overshadow the corresponding ones from the
non-renormalizable term $\bar{H}^cH^c\bar{h}^{\prime}h$.
The resulting $SU(2)_R$ violating mixing of the doublets
in $h$ and $h^{\prime}$ is then unsuppressed and we can
obtain stronger violation of Yukawa unification.

\par
Instead of the pair $\bar\phi$, $\phi$ one can use a pair
of $SU(2)_R$ singlet superfields $\bar{\phi}^{\prime}=
(15,1,1)$, $\phi^{\prime}=(15,1,1)$ with the same PQ and
R charges and the couplings
$\phi^{\prime}\bar{h}^{\prime}h$ and
$\bar{\phi}^{\prime}\phi^{\prime}$. In this case, the
mixing is $SU(2)_R$ invariant and the `asymptotic'
relation $h_t=h_b$ is not violated. The additional
superpotential couplings allowed by the symmetries are:
\begin{equation}
\bar{\phi}^{\prime}(\bar{H}^c)^8,
~\bar{\phi}^{\prime}(H^c)^8,
~\bar{\phi}^{\prime}\bar{H}^cH^c,
~\phi^{\prime}N^2hh^{\prime}.
\label{phiprimecouplings}
\end{equation}
Note the absence of the couplings
$\bar{\phi}^{\prime}(\bar{H}^c)^4$,
$\bar{\phi}^{\prime}(H^c)^4$ which are $SU(2)_R$
triplets. Multiplications of the superpotential terms
by $\phi^{\prime}$, $(\bar{H}^c)^4$, $(H^c)^4$,
$\bar{H}^cH^c$ are generally allowed with the exception
of the multiplication of $S$, $\bar{N}^2N^2$, $N^2h^2$
by a single $\phi^{\prime}$, which is a $SU(4)_c$
15-plet, or a single $\phi^{\prime}(\bar{H}^c)^4$,
$\phi^{\prime}(H^c)^4$, which are $SU(2)_R$
triplets.

\par
Finally, one could introduce both the pairs $\bar\phi$,
$\phi$ and $\bar{\phi}^{\prime}$, $\phi^{\prime}$ at the
same time. The allowed superpotential terms include all
the above mentioned terms and
\begin{equation}
\bar{\phi}^{\prime}\phi(\bar{H}^c)^4,
~\bar{\phi}^{\prime}\phi(H^c)^4,
~\bar{\phi}^{\prime}\phi^2.
\label{extracouplings}
\end{equation}
Multiplications of the superpotential terms by
$\phi$, $\phi^{\prime}$, $(\bar{H}^c)^4$, $(H^c)^4$,
$\bar{H}^cH^c$ are generally allowed with the exception
of the multiplication of $S$, $\bar{N}^2N^2$, $N^2h^2$
by a single factor of $\phi$, $\phi^{\prime}$,
$\phi\phi^{\prime}$, which are $SU(4)_c$ 15-plets, or
a single factor of $\phi^{\prime}(\bar{H}^c)^4$,
$\phi^{\prime}(H^c)^4$. Also, multiplication of
$\bar{\phi}^{\prime}\phi^{\prime}$ by a single $\phi$
is not allowed.

\par
To further analyze the mixing of the doublets in
$h$ and $h^{\prime}$, we must first define properly
the relevant couplings $\bar{h}^{\prime}h^{\prime}$,
$\phi\bar{h}^{\prime}h$ and
$\phi^{\prime}\bar{h}^{\prime}h$. The superfield
$h=(1,2,2)$ is written as
\begin{equation}
h=\left(\begin{array}{cc}h^+_2~,&h^0_1\\ h^0_2~,&h^-_1
\end{array}\right)\equiv\left(\begin{array}{cc}h_2~,
&h_1\end{array}\right).
\label{ewhiggs}
\end{equation}
The color singlet components of the fields
$\bar{h}^{\prime}$, $h^{\prime}$ can be similarly
represented. Under $U_c\in SU(4)_c$, $U_L\in SU(2)_L$
and $U_R\in SU(2)_R$, the relevant fields transforms as:
\begin{eqnarray}
F\rightarrow FU_L^{\dagger}\tilde{U}_c,~
F^c\rightarrow \tilde{U}_c^{\dagger}
\tilde{U}_R^{\dagger}F^c,~
h\rightarrow U_Lh\tilde{U}_R,~
\bar{h}^{\prime}\rightarrow
U_cU_L\bar{h}^{\prime}\tilde{U}_RU_c^{\dagger},~
h^{\prime}\rightarrow
\tilde{U}_c^{\dagger}U_Lh^{\prime}
\tilde{U}_R\tilde{U}_c,~
\nonumber\\
\phi\rightarrow U_cU_R\phi U_R^{\dagger}U_c^{\dagger},~
\phi^{\prime}\rightarrow U_c\phi^{\prime}U_c^{\dagger},
~~~~~~~~~~~~~~~~~~~~~~~~~~~~~~~~~~
\label{transformations}
\end{eqnarray}
where the transpose of a matrix is denoted by tilde.
The Yukawa couplings are $FhF^c$, $Fh^{\prime}F^c$.
From Eq.(\ref{transformations}) and the identities
$\tilde{U}_L\epsilon U_L=\tilde{U}_R\epsilon U_R=
\epsilon$ ($\epsilon$ is the $2\times 2$
antisymmetric matrix with $\epsilon_{12}=1$) which
follow from $\det U_L=\det U_R=1$, one
finds that $\tilde{h}\epsilon\rightarrow U_R
\tilde{h}\epsilon U_L^{\dagger}$, $\bar{h}^{\prime}
\epsilon\rightarrow U_cU_L\bar{h}^{\prime}\epsilon
U_R^{\dagger}U_c^{\dagger}$, $\tilde{h}^{\prime}
\epsilon\rightarrow U_cU_R\tilde{h}^{\prime}\epsilon
U_L^{\dagger}U_c^{\dagger}$. We see that ${\rm tr}
(\bar{h}^{\prime}\epsilon\tilde{h}^{\prime}\epsilon)$,
${\rm tr}(\bar{h}^{\prime}\epsilon\phi\tilde{h}
\epsilon)$, ${\rm tr}(\bar{h}^{\prime}\epsilon
\phi^{\prime}\tilde{h}\epsilon)$ (the traces are taken
with respect to the $SU(4)_c$ and $SU(2)_L$ indices)
are invariant under $G_{PS}$. They correspond to
the `symbolic' couplings $\bar{h}^{\prime}h^{\prime}$,
$\phi\bar{h}^{\prime}h$,
$\phi^{\prime}\bar{h}^{\prime}h$ respectively.

\par
After the breaking of $G_{PS}$ to $G_S$, the fields
$\phi$, $\phi^{\prime}$ acquire vevs $\langle\phi
\rangle$, $\langle\phi^{\prime}\rangle\sim M_{GUT}$.
Substituting them by these vevs in the above
couplings and using Eq.(\ref{ewhiggs}), we obtain
\begin{eqnarray}
{\rm tr}(\bar{h}^{\prime}\epsilon\tilde{h}^{\prime}
\epsilon)=\tilde{\bar{h}}^{\prime}_1\epsilon
h^{\prime}_2+\tilde{h}^{\prime}_1\epsilon
\bar{h}^{\prime}_2+\cdots,
~~~~~~~~~~~~
\label{mass}
\\
{\rm tr}(\bar{h}^{\prime}\epsilon\phi\tilde{h}
\epsilon)=\frac{\langle\phi\rangle}{\sqrt{2}}{\rm tr}
(\bar{h}^{\prime}\epsilon\sigma_3\tilde{h}\epsilon)=
\frac{\langle\phi\rangle}{\sqrt{2}}
(\tilde{\bar{h}}^{\prime}_1\epsilon h_2-\tilde{h}_1
\epsilon\bar{h}^{\prime}_2),
\label{triplet}
\\
{\rm tr}(\bar{h}^{\prime}\epsilon\phi^{\prime}
\tilde{h}\epsilon)=\frac{\langle\phi^{\prime}\rangle}
{\sqrt{2}}{\rm tr}(\bar{h}^{\prime}\epsilon\tilde{h}
\epsilon)=\frac{\langle\phi^{\prime}\rangle}
{\sqrt{2}}(\tilde{\bar{h}}^{\prime}_1\epsilon h_2+
\tilde{h}_1\epsilon\bar{h}^{\prime}_2),
\label{singlet}
\end{eqnarray}
where only the colorless components of
$\bar{h}^{\prime}$ and $h^{\prime}$ are shown in the
right hand side of Eq.(\ref{mass}) and $\sigma_3=
{\rm diag}(1,-1)$. The bilinear terms
between $h_1$, $\bar{h}^{\prime}_1$, $h^{\prime}_1$ and
$h_2$, $\bar{h}^{\prime}_2$, $h^{\prime}_2$ which appear
in Eqs.(\ref{mass}), (\ref{triplet}) and (\ref{singlet})
turn out to be the dominant bilinear terms between these
doublets. Collecting them together, we obtain
\begin{equation}
m\tilde{\bar{h}}^{\prime}_1\epsilon(h^{\prime}_2+
\alpha_2h_2)+m(\tilde{h}^{\prime}_1+\alpha_1
\tilde{h}_1)\epsilon\bar{h}^{\prime}_2,
\label{superheavy}
\end{equation}
where $m$ is the superheavy mass parameter which
multiplies the term in Eq.(\ref{mass}) and
$\alpha_1=(-p\langle\phi\rangle+q\langle\phi^{\prime}
\rangle)/\sqrt{2}m$, $\alpha_2=(p\langle\phi\rangle+
q\langle\phi^{\prime}\rangle)/\sqrt{2}m$ with $p$ and
$q$ being the dimensionless coupling constants which
correspond to the $SU(2)_R$ triplet and singlet
terms in Eqs.(\ref{triplet}) and (\ref{singlet})
respectively. Note that $\alpha_1$, $\alpha_2$ are
in general complex. So, we get two pairs of superheavy
doublets with mass $m$. They are predominantly given
by
\begin{equation}
\bar{h}^{\prime}_1~,~\frac{h^{\prime}_2+\alpha_2h_2}
{(1+|\alpha_2|^2)^{\frac{1}{2}}}~~{\rm and}~~
\frac{h^{\prime}_1+\alpha_1h_1}
{(1+|\alpha_1|^2)^{\frac{1}{2}}}~,~\bar{h}^{\prime}_2.
\label{superdoublets}
\end{equation}
The orthogonal combinations of $h_1$, $h^{\prime}_1$
and $h_2$, $h^{\prime}_2$ constitute the electroweak
doublets:
\begin{equation}
h_1^{ew}=\frac{h_1-\alpha_1^*h^{\prime}_1}
{(1+|\alpha_1|^2)^{\frac{1}{2}}}~~{\rm and}~~
h_2^{ew}=\frac{h_2-\alpha_2^*h^{\prime}_2}
{(1+|\alpha_2|^2)^{\frac{1}{2}}}\cdot
\label{ew}
\end{equation}

\par
We see that, although $h_1$, $h_2$ and $h^{\prime}_1$,
$h^{\prime}_2$ form $SU(2)_R$ doublets, this is, in
general, not true for $h_1^{ew}$, $h_2^{ew}$ since
$\alpha_1$, $\alpha _2$ can be different. However, if
we remove from the model the $SU(2)_R$ triplets
$\bar\phi$, $\phi$ (and, thus, the $SU(2)_R$ triplet
coupling in Eq.(\ref{triplet})), we obtain $\alpha_1
=\alpha _2$ and $h_1^{ew}$, $h_2^{ew}$ do form a
$SU(2)_R$ doublet. The minimal model which can
provide us with an adequate violation of $t-b$ Yukawa
unification for $\mu>0$ is the one including
$\bar\phi$, $\phi$ but not the $SU(2)_R$ singlets
$\bar\phi^{\prime}$, $\phi^{\prime}$. This model,
which we choose to study in detail here, yields
$\alpha_1=-\alpha _2$. Inclusion of both $\bar\phi$,
$\phi$ and $\bar\phi^{\prime}$, $\phi^{\prime}$ can
lead to an arbitrary relation between $\alpha_1$ and
$\alpha _2$ and, thus, introduces an extra complex
parameter. Finally, due to the presence of
$h^{\prime}_1$, $h^{\prime}_2$, the electroweak
doublets are generally non-singlets under $SU(4)_c$
and $b-\tau$ Yukawa unification can also be violated.

\par
The doublets in Eq.(\ref{superdoublets}) must have
zero vevs, which implies that $\langle h_1^{\prime}
\rangle=-\alpha_1\langle h_1\rangle$, $\langle
h_2^{\prime}\rangle=-\alpha_2\langle h_2\rangle$.
Eq.(\ref{ew}) then gives $\langle
h_1^{ew}\rangle=(1+|\alpha_1|^2)^{1/2}\langle h_1
\rangle$, $\langle h_2^{ew}\rangle=(1+|\alpha_2|^2)
^{1/2}\langle h_2\rangle$. From the third generation
Yukawa couplings $y_{33}F_3hF_3^c$,
$2y_{33}^{\prime}F_3h^{\prime}F_3^c$, we obtain
\begin{equation}
m_t=|y_{33}\langle h_2\rangle+y_{33}^{\prime}\langle
h_2^{\prime}\rangle|=|(y_{33}-y_{33}^{\prime}\alpha_2)
\langle h_2\rangle|=\left|\frac{1-\rho\alpha_2/
\sqrt{3}}{(1+|\alpha_2|^2)^{\frac{1}{2}}}y_{33}
\langle h_2^{ew}\rangle\right|,
\label{top}
\end{equation}
where $\rho=y_{33}^{\prime}/y_{33}$ and can be taken
positive by appropriately readjusting the phases of
$h$, $h^{\prime}$. Note that, in the $SU(4)_c$ space,
the doublets in $h^{\prime}$ are proportional
to ${\rm diag}(1/2\sqrt{3},1/2\sqrt{3},1/2\sqrt{3},
-\sqrt{3}/2)$, which is normalized so that the trace
of its square equals unity. Thus, to make
$y_{33}^{\prime}$ directly comparable to $y_{33}$,
we included a factor of two in defining the
corresponding Yukawa coupling. The masses $m_b$,
$m_\tau$ are, similarly, found:
\begin{equation}
m_b=\left|\frac{1-\rho\alpha_1/
\sqrt{3}}{(1+|\alpha_1|^2)^{\frac{1}{2}}}y_{33}
\langle h_1^{ew}\rangle\right|,~
m_\tau=\left|\frac{1+\sqrt{3}\rho\alpha_1}
{(1+|\alpha_1|^2)^{\frac{1}{2}}}y_{33}
\langle h_1^{ew}\rangle\right|\cdot
\label{bottomtau}
\end{equation}
From Eqs.(\ref{top}) and (\ref{bottomtau}), we see
that the exact equality of the `asymptotic' Yukawa
couplings $h_t$, $h_b$, $h_\tau$ is now replaced by
the quasi-unification condition:
\begin{equation}
h_t:h_b:h_\tau=
\left|\frac{1-\rho\alpha_2/\sqrt{3}}
{(1+|\alpha_2|^2)^{\frac{1}{2}}}\right|:
\left|\frac{1-\rho\alpha_1/\sqrt{3}}
{(1+|\alpha_1|^2)^{\frac{1}{2}}}\right|:
\left|\frac{1+\sqrt{3}\rho\alpha_1}
{(1+|\alpha_1|^2)^{\frac{1}{2}}}\right|\cdot
\label{quasi}
\end{equation}
This condition depends on two complex ($\alpha_1$,
$\alpha_2$) and one real ($\rho>0$) parameter.

\par
Note that the mixing between $h_1$, $h_1^{\prime}$
(i.e., $\alpha_1\neq 0$) and the fact that the Higgs
superfield $h^{\prime}$ possesses Yukawa couplings to
the `matter' superfields (i.e., $\rho\neq 0$) are
crucial for violating $b-\tau$ Yukawa unification,
which is though not affected by the mixing between
$h_2$, $h_2^{\prime}$ (i.e., the value of $\alpha_2$).
On the contrary, violation of $t-b$ Yukawa unification
requires that $\alpha_1\neq\alpha_2$, which can be
achieved only in the presence of a $SU(2)_R$ triplet
bilinear term between $\bar{h}^{\prime}$ and $h$. In
summary, the minimal requirements for full violation
of Yukawa unification are $\rho,~\alpha_1\neq 0$,
$\alpha_1\neq\alpha_2$. In the minimal model (with
$\bar{\phi}$, $\phi$ but not $\bar{\phi}^{\prime}$,
$\phi^{\prime}$) which we will study here, $\alpha_1
=-\alpha_2$ and, thus, Eq.(\ref{quasi}) takes the
simple form
\begin{equation}
h_t:h_b:h_\tau=|1+c|:|1-c|:|1+3c|,
\label{minimal}
\end{equation}
where $c=\rho\alpha_1/\sqrt{3}$. This `asymptotic'
relation depends on a single complex parameter. For
simplicity, we will restrict our analysis to real
values of $c$ only.

\par
For completeness, we also give the `asymptotic'
relation between the Yukawa couplings in the model
without $\bar{\phi}$, $\phi$, $\bar{\phi}^{\prime}$,
$\phi^{\prime}$, which, although not suitable for
$\mu>0$, may be adequate for $\mu<0$. Under $G_{PS}$,
the superfields $\bar{H}^c$, $H^c$ transform as:
$\bar{H}^c\rightarrow\bar{H}^c\tilde{U}_R\tilde{U}_c$,
$H^c\rightarrow\tilde{U}_c^{\dagger}
\tilde{U}_R^{\dagger}H^c$. Thus, $\tilde{\bar{H}}^c
\tilde{H}^c\rightarrow U_cU_R\tilde{\bar{H}}^c
\tilde{H}^cU_R^{\dagger}U_c^{\dagger}$, and ${\rm tr}
(\bar{h}^{\prime}\epsilon\tilde{\bar{H}}^c
\tilde{H}^c\tilde{h}\epsilon)$ is invariant under
$G_{PS}$ containing both the $SU(2)_R$ singlet and
triplet `symbolic' couplings $\bar{H}^cH^c
\bar{h}^{\prime}h$. In the $SU(2)_R$ space, $\langle
\tilde{\bar{H}}^c\rangle\langle\tilde{H}^c\rangle=
{\rm diag}(v_0^2,0)=v_0^2\sigma_0/2+v_0^2\sigma_3/2$,
where $v_0=M_{GUT}/g_{GUT}$, with $g_{GUT}$ being the
GUT gauge coupling constant, and $\sigma_0$ is the
unit $2\times 2$ matrix. Moreover, in the $SU(4)_c$
space, $\langle\tilde{\bar{H}}^c\rangle\langle
\tilde{H}^c\rangle={\rm diag}(0,0,0,v_0^2)$. So,
replacing $\bar{H}^c$, $H^c$ by their vevs in
the singlet and triplet couplings in ${\rm tr}
(\bar{h}^{\prime}\epsilon\tilde{\bar{H}}^c
\tilde{H}^c\tilde{h}\epsilon)$ (with dimensionless
coefficients equal to $q^{\prime}$ and $p^{\prime}$),
we obtain $(-\sqrt{3}v_0^2/4){\rm tr}(\bar{h}^{\prime}
\epsilon\tilde{h}\epsilon)$ and $(-\sqrt{3}v_0^2/4)
{\rm tr}(\bar{h}^{\prime}\epsilon\sigma_3\tilde{h}
\epsilon)$. We, thus, again end up with
Eq.(\ref{quasi}) but with $\alpha_1=-\sqrt{3}v_0^2
(q^{\prime}-p^{\prime})/4mM_S$, $\alpha_2=-\sqrt{3}
v_0^2(q^{\prime}+p^{\prime})/4mM_S$, which are
suppressed by $M_{GUT}/M_S$.

\section{The Resulting MSSM}
\label{mssm}

\par
We will now concentrate on the minimal model which
includes $\bar{\phi}$, $\phi$ but not
$\bar{\phi}^{\prime}$, $\phi^{\prime}$. This model,
below $M_{GUT}$, reduces to the MSSM supplemented by
the `asymptotic' Yukawa coupling quasi-unification
condition in Eq.(\ref{minimal}), where $c$ is taken
real for simplicity. We will assume universal soft
SUSY breaking terms at $M_{GUT}$, i.e., a common mass
for all scalar fields $m_0$, a common gaugino mass
$M_{1/2}$ and a common trilinear scalar coupling
$A_0$. So the resulting MSSM is actually the CMSSM.
Furthermore, we will concentrate on the $\mu>0$
case for reasons which we already explained.

\par
We will closely follow the notation as well as the
RG and radiative electroweak breaking analysis of
Ref.\cite{cdm} for the CMSSM with the improvements
of Ref.\cite{cd2} (recall that the sign of
$\mu$ in these references is opposite to the one
adopted here). These improvements include the
employment of the full one-loop
corrections to the effective potential for the
electroweak breaking and to certain particle
masses taken from Ref.\cite{pierce}. They also
include the incorporation of the two-loop
corrections to the CP-even neutral Higgs boson
mass matrix by using {\tt FeynHiggsFast} \cite{fh}
and the introduction \cite{drees} of a variable
common SUSY threshold $M_{SUSY}=(m_{\tilde t_1}
m_{\tilde t_2})^{1/2}$ ($\tilde t_{1,2}$ are the
stop mass eigenstates) where the effective potential
is minimized, $m_A$ (the CP-odd Higgs boson mass)
and $\mu$ are evaluated, and the MSSM RG equations
are replaced by the SM ones. Note that, at
$M_{SUSY}$, the size of the one-loop corrections to
$m_A$ and $\mu$ is \cite{drees} minimal and, thus,
the accuracy in the determination of these
quantities is maximal.

\par
For any given $m_b(M_Z)$ in its $95\%$ c.l.
range ($2.684-3.092~{\rm GeV}$), we can determine
the parameters $c$ and $\tan\beta$ at
$M_{SUSY}$ so that the `asymptotic' condition in
Eq.(\ref{minimal}) is satisfied. We use fixed
values for the running top quark mass
$m_t(m_t)=166~{\rm GeV}$ and the running tau lepton
mass $m_\tau(M_Z)=1.746~{\rm{GeV}}$ and incorporate
not only the SUSY correction to the bottom quark
mass but also the SUSY threshold correction to
$m_\tau(M_{SUSY})$ from the approximate formula of
Ref.\cite{pierce}. This correction arises mainly
from chargino/tau sneutrino ($\tilde\nu_{\tau}$)
loops and, for $\mu>0$, leads \cite{cd2} to a small
reduction of $\tan\beta$.

\par
After imposing the conditions of gauge coupling
unification, successful electroweak breaking and
Yukawa quasi-unification in Eq.(\ref{minimal}),
we are left with three free input parameters $m_0$,
$M_{1/2}$ and $A_0$ (see Refs.\cite{cd2,cdm}). In
order to make the notation physically more
transparent, we replace $m_0$ and $M_{1/2}$
equivalently by the mass $m_{LSP}$ (or
$m_{\tilde\chi}$) of the lightest supersymmetric
particle (LSP), which turns out to be the lightest
neutralino ($\tilde\chi$), and the relative mass
splitting  $\Delta_{\tilde\tau_2}=(m_{\tilde\tau_2}
-m_{\tilde\chi})/m_{\tilde\chi}$ between the lightest
stau mass eigenstate ($\tilde\tau_2$) and the LSP.
We will study the parameter space of this model
which is compatible with all the available
phenomenological and cosmological constraints.

\par
An important constraint is obtained by considering the
inclusive branching ratio ${\rm BR}(b\rightarrow s
\gamma)$ of the decay process $b\rightarrow s\gamma$.
The present best average of this branching ratio, which
is found from the available experimental data \cite{cleo},
is $3.24\times 10^{-4}$ with an experimental error of
$\pm 0.38\times 10^{-4}$ and an asymmetric `theoretical'
error due to model dependence of $[+0.26,-0.23]\times
10^{-4}$ (assuming no correlation between the
experimental systematics, which should not be too far
from reality).

\par
We calculate ${\rm BR}(b\rightarrow s\gamma)$ using the
formalism of Ref.\cite{kagan}, where the SM contribution
is factorized out. This contribution includes the
next-to-leading order (NLO) QCD and the leading order (LO)
QED corrections. The charged Higgs boson contribution to
${\rm BR}(b\rightarrow s\gamma)$ is evaluated by including
the NLO QCD corrections from Ref.\cite{nlohiggs}. The
dominant SUSY contribution includes the NLO QCD
corrections from Ref.\cite{nlosusy}, which hold for large
$\tan\beta$ (see also Ref.\cite{carena}). The error in the
SM contribution to ${\rm BR}(b\rightarrow s\gamma)$ is
made of two components. The first is due to the
propagation of the experimental errors in the input
parameters and is \cite{gambino} about $\pm 0.23\times
10^{-4}$. The second originates from the dependence on the
renormalization and matching scales and is \cite{gerardo}
about $\pm 6\%$ if the NLO QCD corrections are included.
These errors remain \cite{gerardo} basically the same even
after including the charged Higgs and SUSY contributions
corrected at NLO.

\par
In order to construct the $95\%$ c.l. range of
${\rm BR}(b\rightarrow s\gamma)$, we first add in
quadrature the experimental error
($\pm 0.38\times 10^{-4}$) and the error originating
from the input parameters ($\pm 0.23\times 10^{-4}$).
This yields the overall standard deviation ($\sigma$).
The asymmetric error coming from model dependence
($[+0.26,-0.23]\times 10^{-4}$) and the error from
scale dependence ($6\%$) are then added linearly
\cite{ganis} on both ends of the $\pm 2-\sigma$ range.
For simplicity, we take a constant error from scale
dependence, which we evaluate at the central experimental
value of ${\rm BR}(b\rightarrow s\gamma)$. The $95\%$ c.l.
range of this branching ratio then turns out to be about
$(1.9-4.6)\times 10^{-4}$.

\par
The recent measurement \cite{muon} of the anomalous magnetic
moment of the muon $a_\mu\equiv (g_\mu-2)/2$ provides an
additional significant constraint. The deviation of $a_\mu$
from its predicted value in the SM, $\delta a_\mu$, is
found to lie, at $95\%$ c.l., in the range from $-6\times
10^{-10}$ to $58\times 10^{-10}$. This range is derived
using the calculations (see e.g., Ref.\cite{marciano}) of
$a_\mu$ in the SM which are based on the evaluation of the
hadronic vacuum polarization contribution of
Ref.\cite{davier}. However, we take here the recently
corrected \cite{sign} sign of the pseudoscalar pole
contribution to the light-by-light scattering correction
to $a_\mu$. This corrected sign reduces considerably the
discrepancy between the SM and the measured value of
$a_\mu$. It also relaxes the restrictions on the parameter
space of the CMSSM. In particular, the $\mu<0$ case, which
leads to negative $\delta a_\mu$, is no longer disfavored.
However, the sparticles, in this case, cannot be as light
as in the $\mu>0$ case, where $\delta a_\mu>0$. The
calculation of $\delta a_\mu$ in the CMSSM is performed
here by using the analysis of Ref.\cite{gminus2}.

\par
Another constraint results from the requirement that the
relic abundance $\Omega_{LSP}~h^2$ of the LSP in the
universe does not exceed the upper limit on the cold dark
matter (CDM) abundance which is derived from observations
($\Omega_{LSP}$ is the present energy density of the LSP
over the critical energy density of the universe and $h$
is the present value of the Hubble parameter in units of
$100~\rm{km}~\rm{sec}^{-1}~\rm{Mpc}^{-1}$). From the
recent results of DASI \cite{dasi}, one finds that the
$95\%$ c.l. range of $\Omega_{CDM}~h^{2}$ is $0.06-0.22$.
Therefore, we require that $\Omega_{LSP}~h^2$ does
not exceed 0.22.

\par
Here, the LSP ($\tilde\chi$) is an almost pure bino.
Its relic abundance will be calculated by
{\tt micrOMEGAs} \cite{micro}, which is the most
complete code available. (A similar calculation has
appeared in Ref.\cite{baer}.) It includes all the
coannihilations \cite{coan} of neutralinos,
charginos, sleptons, squarks and gluinos. The
exact tree-level cross sections are used and
are accurately thermally averaged. Also, poles
and thresholds are properly handled and one-loop
QCD corrected Higgs decay widths \cite{width}
are used, which is the main improvement
provided by Ref.\cite{micro}. The SUSY
corrections \cite{susy} to these widths are,
however, not included. Fortunately, in our case,
their effect is much smaller than that of the
QCD corrections.

\par
In order to have an independent check of
{\tt micrOMEGAs}, we also use the following alternative
method for calculating $\Omega_{LSP}~h^2$ in
our model. In most of the parameter space where
coannihilations are unimportant, $\Omega_{LSP}~h^2$
can be calculated by using {\tt DarkSUSY}
\cite{darksusy}. This code employs the complete
tree-level (and in some cases one-loop) cross sections
of the relevant processes with properly treating all
resonances and thresholds and performs accurate
numerical integration of the Boltzmann equation.
Its neutralino annihilation part is in excellent
numerical agreement with the recent exact analytic
calculation of Ref.\cite{roberto}. Moreover, it
includes all coannihilations between neutralinos and
charginos which are, however, insignificant for our
model. Its main defect is that it uses the tree-level
Higgs decay widths. This can be approximately corrected
if, in evaluating the Higgs decay widths, we replace
$m_b(m_b)$ by $m_b$ at the mass of the appropriate
Higgs boson in the couplings of the $b$-quark to the
Higgs bosons (see Ref.\cite{micro}).

\par
In the region of the parameter space where
coannihilations come into play, the next-to-lightest
supersymmetric particle (NLSP) turns out to be the
$\tilde\tau_2$ and the only relevant coannihilations
are the bino-stau ones \cite{cdm,ellis}. In this
region, which is given by
$\Delta_{\tilde\tau_2}<0.25$, we calculate
$\Omega_{LSP}~h^2$ by using an improved version
of the analysis of Ref.\cite{cdm} (the sign of $\mu$
in this reference is opposite to the one adopted here).
This analysis, which has been applied in
Refs.\cite{cd2,hw}, includes bino-stau coannihilations
for all $\tan\beta$'s (see also
Refs.\cite{cmssm,arnowitt}) and is based on a series
expansion of the thermally averaged cross sections in
$x_F^{-1}=T_F/m_{\tilde\chi}$, with $T_F$ being the
freeze-out temperature. This expansion is, however,
inadequate when we have to treat resonances, and these
are crucial for bino annihilation. So, we need to
improve the bino annihilation part in Ref.\cite{cdm}.
This can be achieved by evaluating the corresponding
expansion coefficients $a_{\tilde\chi\tilde\chi}$ and
$b_{\tilde\chi\tilde\chi}$ not as in this reference
but as follows. From {\tt DarkSUSY}, we find the
values of $\Omega_{\tilde\chi}~h^2$ and $x_F$ which
correspond to essentially just bino annihilation.
Using the appropriate formulas of Ref.\cite{cdm}, we
then extract the `improved' expansion coefficients
$a_{\tilde\chi\tilde\chi}$ and
$b_{\tilde\chi\tilde\chi}$.

\par
There is one more improvement which we need to do
in Ref.\cite{cdm} (and was already used in
Ref.\cite{vergados}) to make it applicable to the
present case. The cross sections of the processes
$\tilde\tau_2\tilde\tau_2^\ast\rightarrow W^+ W^-
,\ H^+ H^-$ were calculated without including the
tree-graphs with $\tilde\nu_{\tau}$ exchange in
the t-channel \cite{falk}. Thus, the contribution
to $a_{\tilde\tau_2\tilde\tau_2^\ast}$ from the
process $\tilde\tau_2\tilde\tau_2^\ast\rightarrow
W^+ W^-$ appearing in Table II of Ref.\cite{cdm}
should be corrected by adding to it
$$\frac{g_{\tilde\tau_2\tilde\nu_{\tau}W^\pm}^2
(1-\hat m_W^2)^{1/2}}{4\pi m_W^4(1-\hat m_W^2+
\hat m_{\tilde\nu_{\tau}}^2)}\left[
2g_{\tilde\tau_2\tilde\nu_{\tau}W^\pm}^2
\frac{(1-\hat m_W^2)^2m_{\tilde\tau_2}^2}
{1-\hat m_W^2+\hat m_{\tilde\nu_{\tau}}^2}\right.
$$
\begin{equation}
\left. -(2-3\hat m_W^2+\hat m_W^4)\left(
\frac{g_Hg_{HW^+W^-}}{\hat m_H^2 -4}+
\frac{g_hg_{hW^+ W^-}}{\hat m_h^2-4}+
g_{\tilde\tau_2 \tilde\tau_2 W^+ W^-}
m_{\tilde\tau_2}^2\right)\right],
\label{WW}
\end{equation}
with $g_{\tilde\tau_2\tilde\nu_{\tau}W^\pm}=gs_\theta/
\sqrt{2}$. Also, the contribution to
$a_{\tilde\tau_2\tilde\tau_2^\ast}$ from
$\tilde\tau_2\tilde\tau_2^\ast\rightarrow H^+ H^-$ is now
given by Eq.(26) of Ref.\cite{cdm} with the expression in
the last parenthesis corrected by adding to it the
quantity $g_{\tilde\tau_2\tilde\nu_{\tau} H^\pm}^2/
(\hat m_{H^{\pm}}^2-\hat m_{\tilde\nu_{\tau}}^2-1)$,
where
$$g_{\tilde\tau_2\tilde\nu_{\tau}H^\pm}=g[(m_W^2
\sin{2\beta}-m_\tau^2\tan\beta)s_\theta+(A_\tau\tan\beta+
\mu)m_\tau c_\theta]/\sqrt{2}m_W$$
(with the present sign convention for $\mu$). The
resulting correction to $\Omega_{\tilde\chi}~h^2$, in the
case of the process $\tilde\tau_2\tilde\tau_2^\ast
\rightarrow W^+ W^-$, varies from about $8\%$ to about
$1\%$ as $\Delta_{\tilde\tau_2}$ increases from 0 to
0.25. On the contrary, the correction from $\tilde\tau_2
\tilde\tau_2^\ast\rightarrow H^+ H^-$ is negligible.

\par
We find that the alternative method for calculating the
neutralino relic abundance in our model, which we have just
described, yields results which are in excellent agreement
with {\tt micrOMEGAs}. In practice, however, we use the
code {\tt micrOMEGAs} since it has the extra advantage of
being much faster.

\par
We will also impose the $95\%$ c.l. LEP bound on the
lightest CP-even neutral Higgs boson mass
$m_h>114.1~{\rm GeV}$ \cite{higgs}. In the CMSSM, this
bound holds almost always for all $\tan\beta$'s, at least
as long as CP is conserved. Finally, for the values of
$\tan\beta$ which appear here (about 60), the CDF results
yield the $95\%$ c.l. bound $m_A>110~{\rm GeV}$ \cite{cdf}.

\section{The Allowed Parameter Space}
\label{parameters}

\par
We now proceed to the derivation of the restrictions
which are imposed by the various phenomenological and
cosmological constraints presented in Sec.\ref{mssm}
on the parameters of the CMSSM (with $\mu>0$)
supplemented with the Yukawa quasi-unification
condition in Eq.(\ref{minimal}) (with $0<c<1$). The
restrictions on the $m_{LSP}-\Delta_{\tilde\tau_2}$
plane, for $A_0=0$ and with the central value of
$\alpha_s(M_Z)=0.1185$, are shown in Fig.\Figa . The
dashed (dotted) lines correspond to the $95\%$ c.l.
lower (upper) experimental bound on $m_b(M_Z)$ which
is $2.684~{\rm GeV}$ ($3.092~{\rm GeV}$)
(see Sec.\ref{intro}), while the solid lines
correspond to the central experimental value of
$m_b(M_Z)=2.888~{\rm GeV}$. We will follow this
convention in the subsequent four figures too. From
left to right, the dashed (dotted) lines depict the
$95\%$ c.l. lower bounds on $m_{LSP}$ from the
constraints $m_A>110~{\rm GeV}$,
${\rm BR}(b\rightarrow s\gamma)>1.9\times 10^{-4}$
and $\delta a_\mu<58\times 10^{-10}$, and the $95\%$
c.l. upper bound on $m_{LSP}$ from
$\Omega_{LSP}~h^2<0.22$. The constraints
${\rm BR}(b\rightarrow s\gamma)<4.6\times 10^{-4}$
and $\delta a_\mu>-6\times 10^{-10}$ do not restrict
the parameters since they are always satisfied for
$\mu>0$. The left solid line depicts the lower bound
on $m_{LSP}$ from $m_h>114.1~{\rm GeV}$ which does not
depend much on $m_b(M_Z)$, while the right solid line
corresponds to $\Omega_{LSP}~h^2=0.22$ for the central
value of $m_b(M_Z)$.

\par
We observe that the lower bounds on $m_{LSP}$ are
generally not so sensitive to the variations of
$m_b(M_Z)$ within its $95\%$ c.l. range. The most
sensitive of these bounds is the one from
$m_A>110~{\rm GeV}$, while the bound from
$m_h>114.1~{\rm GeV}$ is practically
$m_b(M_Z)$-independent. Actually, this bound
overshadows all the other lower bounds for all
$\Delta_{\tilde\tau_2}$'s and, being essentially
$\Delta_{\tilde\tau_2}$-independent too, sets, for
$A_0=0$ and $\alpha_s(M_Z)=0.1185$, an overall
constant $95\%$ c.l. lower bound on $m_{LSP}$ of
about $138~{\rm GeV}$. Contrary to the lower bounds,
the line from the LSP relic abundance is extremely
sensitive to the value of $m_b(M_Z)$. In particular,
its almost vertical part is considerably displaced
to higher $m_{LSP}$'s as $m_b(M_Z)$ decreases. We
will explain this behavior later.

\par
In Fig.\Figb, we depict $m_A$ and $M_{SUSY}$
versus $m_{LSP}$ for various
$\Delta_{\tilde\tau_2}$'s, $A_0=0$ and the
central values of $m_b(M_Z)$ and $\alpha_s(M_Z)$.
We see that $m_A$ is always smaller than
$2m_{LSP}$ but close to it. Thus, generally, the
neutralino annihilation via the s-channel exchange
of an $A$-boson is by far the dominant
(co)annihilation process. We also observe that,
as $m_{LSP}$ or $\Delta_{\tilde\tau_2}$ increase,
we move away from the $A$-pole, which thus becomes
less efficient. As a consequence,
$\Omega_{LSP}~h^2$ increases with $m_{LSP}$ or
$\Delta_{\tilde\tau_2}$ (see Fig.\Figc).

\par
In Fig.\Figc, we show $\Omega_{LSP}~h^2$
as a function of $m_{LSP}$ for various
$\Delta_{\tilde\tau_2}$'s, $A_0=0$,
$m_b(M_Z)=2.888~{\rm GeV}$ and
$\alpha_s(M_Z)=0.1185$. The solid lines are
obtained by using {\tt micrOMEGAs}.
The crosses are from the alternative
method described in Sec.\ref{mssm}, which
combines {\tt DarkSUSY} with the bino-stau
coannihilation calculation of Ref.\cite{cdm}.
In order to mimic, in {\tt DarkSUSY}, the effect
of the one-loop QCD corrections to the Higgs
decay widths, we can generally use the $b$-quark
mass (including the SUSY corrections) evaluated
at $m_A$ since the $A$-pole is by far
the most important. As it turns out, $m_b(m_A)$
depends very weakly on the values of the input
parameters $m_{LSP}$ and $\Delta_{\tilde\tau_2}$,
and thus the uniform use of a constant mean
value of $m_b(m_A)$ should be adequate. We find
that excellent agreement between {\tt micrOMEGAs}
and our alternative method is achieve if, in
{\tt DarkSUSY}, we use a constant default value
for the $b$-quark mass which is equal to about
$2.5~{\rm GeV}$ (see Fig.\Figc).

\par
The importance for our calculation of the
one-loop QCD corrections to the Higgs
decay widths can be easily concluded from
Fig.\Figd, where we show $\Omega_{LSP}~h^2$
versus $m_{LSP}$ for $\Delta_{\tilde\tau_2}=1$,
$A_0=0$ and the central values of
$m_b(M_Z)$, $\alpha_s(M_Z)$. The LSP relic
abundance is calculated by {\tt micrOMEGAs} with
(thick solid line) or without (faint solid line)
the inclusion of the one-loop QCD corrections
to the Higgs widths. We see that the results
differ appreciably. For comparison, we also draw
$\Omega_{LSP}~h^2$ calculated by our alternative
method with (thick crosses) or without (faint
crosses) the above `correction' of the $b$-quark
mass. The agreement with {\tt micrOMEGAs} is
really impressive in both cases.

\par
In Fig.\Fige, we present $m_A$ and
$M_{SUSY}$ as functions of $m_{LSP}$ for
$\Delta_{\tilde\tau_2}=1$, $A_0=0$ and with the
lower, upper and central values of $m_b(M_Z)$ for
$\alpha_s(M_Z)=0.1185$. We see that $m_A$
increases and approaches $2m_{LSP}$ as $m_b(M_Z)$
decreases. As a consequence, the effect of the
$A$-pole is considerably enhanced and
$\Omega_{LSP}~h^2$ is reduced. This explains the
large displacement of the vertical part of the
$\Omega_{LSP}~h^2=0.22$ line in Fig.\Figa\ at low
$m_b(M_Z)$'s.

\par
As $\Delta_{\tilde\tau_2}$ becomes small and
approaches zero, bino-stau coannihilations become
very important and strongly dominate over the pole
annihilation of neutralinos. This leads to a very
pronounced reduction of the neutralino relic
abundance and, consequently, to a drastic increase
of the upper bound on $m_{LSP}$. We, thus, obtain
the almost horizontal parts of the
$\Omega_{LSP}~h^2=0.22$ lines in Fig.\Figa.

\par
The shaded area depicted in Fig.\Figa, which lies
between the $m_h=114.1~{\rm GeV}$ line and the
$\Omega_{LSP}~h^2=0.22$ line corresponding to
the central value of $m_b(M_Z)$, constitutes the
allowed range in the $m_{LSP}-
\Delta_{\tilde\tau_2}$ plane for $A_0=0$,
$m_b(M_Z)=2.888~{\rm GeV}$ and
$\alpha_s(M_Z)=0.1185$. For practical reasons, we
limited our investigation to
$\Delta_{\tilde\tau_2}\leq 2$ and
$m_{LSP}\leq 500~{\rm GeV}$. We see that
allowing $m_b(M_Z)$ to vary within its $95\%$
c.l. experimental range considerably enlarges
the allowed area in Fig.\Figa. On the other hand,
fixing $m_b(M_Z)$ to its upper bound yields the
smallest possible allowed area for $A_0=0$ and
$\alpha_s(M_Z)=0.1185$, which is, however, still
quite wide. In any case, we conclude that, for
$A_0=0$, there exists a natural and wide parameter
space which is consistent with all the available
phenomenological and cosmological requirements.

\par
In the allowed (shaded) area of Fig.\Figa\ which
corresponds to the central value of $m_b(M_Z)$,
the parameter $c$ ($\tan\beta$) varies between
about 0.15 and 0.20 (58 and 59). For $m_b(M_Z)$
fixed to its lower or upper bound, we find that,
in the corresponding allowed area, the parameter
$c$ ($\tan\beta$) ranges between about 0.17 and
0.23 (59 and 61) or 0.13 and 0.17 (56 and 58).
We observe that, as we increase $m_b(M_Z)$, the
parameter $c$ decreases and we get closer to
exact Yukawa unification. This behavior is
certainly consistent with the fact that the
value of $m_b(M_Z)$ which corresponds to exact
Yukawa unification lies well above its $95\%$
c.l. range. The LSP mass is restricted to be
higher than about $138~{\rm GeV}$ for $A_0=0$
and $\alpha_s(M_Z)=0.1185$, with the minimum being
practically $\Delta_{\tilde\tau_2}$-independent.
At this minimum, $c\approx 0.16-0.20$
($c\approx 0.13-0.23$) and
$\tan\beta\approx 59$
($\tan\beta\approx 58-61$) for
$m_b(M_Z)=2.888~{\rm GeV}$
($m_b(M_Z)=2.684-3.092~{\rm GeV}$).

\par
The `asymptotic' Yukawa quasi-unification
condition in Eq.(\ref{minimal}) with $0<c<1$
yields
\begin{equation}
\delta h\equiv-\frac{h_b-h_t}{h_t}=
\frac{h_\tau-h_t}{h_t}=\frac{2c}{1+c}\cdot
\label{splitting}
\end{equation}
This means that the bottom and tau Yukawa
couplings split from the top Yukawa coupling by
the same amount but in opposite directions, with
$h_b$ becoming smaller than $h_t$. For $A_0=0$
and with the central values of $m_b(M_Z)$ and
$\alpha_s(M_Z)$, the splitting $\delta h$ ranges
from about 0.26 to 0.33. Allowing $m_b(M_Z)$ to
vary in its $95\%$ c.l. range, however, we obtain
a larger range for $\delta h$ which is about
$0.22-0.38$. At the minimum of
$m_{LSP}\approx 138~{\rm GeV}$ ,
$\delta h\approx 0.28-0.33$
($\delta h\approx 0.23-0.38$) for
$m_b(M_Z)=2.888~{\rm GeV}$
($m_b(M_Z)=2.684-3.092~{\rm GeV}$).

\par
For simplicity, $\alpha_s(M_Z)$ was fixed to its
central experimental value (equal to 0.1185)
throughout our calculation. We could, however,
let it vary in its $95\%$ c.l. experimental range
$0.1145-0.1225$, which would significantly widen
the $95\%$ c.l. range of $m_b(M_Z)$ to
$2.616-3.157~{\rm GeV}$, with the lower (upper)
bound corresponding to the upper (lower) bound on
$\alpha_s(M_Z)$. This would lead to a sizable
enlargement of the allowed area due to the
sensitivity of the LSP relic abundance to the
$b$-quark mass.

\par
In Figs.\Figf, \Figg\ and \Figh, we
present the restrictions on the
$m_{LSP}-A_0/M_{1/2}$ plane for
$m_b(M_Z)=2.888~{\rm GeV}$,
$\alpha_s(M_Z)=0.1185$,
and $\Delta_{\tilde\tau_2}=0,~1~{\rm and}~2$
respectively. We use solid, dashed, dot-dashed
and dotted lines to depict the lower bounds on
$m_{LSP}$ from $m_A>110~{\rm GeV}$,
${\rm BR}(b\rightarrow s\gamma)>1.9\times 10^{-4}$,
$\delta a_\mu<58\times 10^{-10}$ and
$m_h>114.1~{\rm GeV}$ respectively. The upper
bounds on $m_{LSP}$ from $\Omega_{LSP}~h^2<0.22$
are represented by double dot-dashed lines.
Note that, in Fig.\Figf, the upper bound on
$m_{LSP}$ does not appear since it lies at
$m_{LSP}>500~{\rm GeV}$. This is due to the fact
that, since the bino and stau masses are
degenerate in this case, the bino-stau coannihilation
is considerably enhanced leading to a strong
reduction of the LSP relic abundance. In order to
illustrate the quick reduction of $\Omega_{LSP}~h^2$
as $\Delta_{\tilde\tau_2}$ approaches zero, we
display in Fig.\Figf\ the bounds from
$\Omega_{LSP}~h^2<0.22$ (double dot-dashed lines) for
$\Delta_{\tilde\tau_2}=0.1~{\rm and}~0.03$ too. The
reduction of $\Omega_{LSP}~h^2$ by bino-stau
coannihilation is also the reason for the fact that
the allowed (shaded) areas in Figs.\Figf, \Figg\ and
\Figh\ are more sizable for smaller
$\Delta_{\tilde\tau_2}$'s. The lower bound on
$m_{LSP}$ overshadows all the other lower bounds in
all cases and is only very mildly dependent on
$\Delta_{\tilde\tau_2}$.

\par
In Figs.\Figi\ and \Figj, we present the
restrictions on the $m_{LSP}-\Delta_{\tilde\tau_2}$
plane for $m_b(M_Z)=2.888~{\rm GeV}$,
$\alpha_s(M_Z)=0.1185$, and
$A_0/M_{1/2}=3~{\rm and}~-3$ respectively. We
follow the same notation for the various bounds on
$m_{LSP}$ and the allowed areas as in Figs.\Figf,
\Figg\ and \Figh. We observe that, for
$A_0/M_{1/2}=3$ ($A_0/M_{1/2}=-3$), the lower bound
on $m_{LSP}$ from the constraint
${\rm BR}(b\rightarrow s\gamma)>1.9\times 10^{-4}$
is overshadowed by (overshadows) the bounds from
the requirements $m_A>110~{\rm GeV}$ and
$\delta a_\mu<58\times 10^{-10}$. In all cases,
however, the overall lower bound on $m_{LSP}$ is set
by $m_h>114.1~{\rm GeV}$.

\par
For presentation purposes, we restricted our
analysis to $m_{LSP}\leq 500~{\rm GeV}$,
$\Delta_{\tilde\tau_2}\leq 2$,
$-3\leq A_0/M_{1/2}\leq 3$. The allowed ranges
in Figs.\Figa, \Figf, \Figg, \Figh, \Figi\ and
\Figj\ constitute sections or boundaries of the
part of the overall allowed parameter space for
$m_b(M_Z)=2.888~{\rm GeV}$ and
$\alpha_s(M_Z)=0.1185$ which is contained in
the investigated range of parameters. We found
that, within this part of the overall allowed
parameter space, $\tan\beta$ ranges between
about 58 and 61. Also, the parameter $c$
ranges between about 0.15 and 0.21 and, thus,
the `asymptotic' splitting between the bottom
(or tau) and the top Yukawa couplings varies in
the range $26-35\%$. Moreover, the smallest
value of $m_{LSP}$ is about $107~{\rm GeV}$ and
is achieved at $A_0/M_{1/2}=-3$ and practically
any $\Delta_{\tilde\tau_2}$. The corresponding
$\delta h\approx 0.30-0.35$ and
$\tan\beta\approx 60-61$. Allowing $m_b(M_Z)$
and $\alpha_s(M_Z)$ to vary within their $95\%$
c.l. ranges will further widen the overall
allowed parameter space within the investigated
range of parameters as well as the allowed
ranges of the parameters $c$ (or $\delta h$) and
$\tan\beta$. The minimal value of $m_{LSP}$ is,
though, not much affected by varying $m_b(M_Z)$.
Variations of $\alpha_s(M_Z)$, however, lead to
small ($\sim 10~{\rm GeV}$) fluctuations of the
minimal $m_{LSP}$.

\par
We see that the required deviation from Yukawa
unification, for $\mu>0$, is not so small. In
spite of this, the restrictions from Yukawa
unification are not completely lost but only
somewhat weakened. In particular, $\tan\beta$
remains large and close to 60. Actually, our
model is much closer to Yukawa unification than
generic models where the Yukawa couplings can
differ even by orders of magnitude. Also, the
deviation from Yukawa unification is generated
here in a natural, systematic, controlled and
well-motivated way. Finally, recall that the
required size of the violation of Yukawa
unification forced us to extend the initial
model of Sec.\ref{model} which could only
provide a suppressed violation.

\section{conclusions}
\label{conclusions}

\par
We constructed a class of concrete SUSY GUTs based
on the PS gauge group which naturally lead
to a moderate violation of `asymptotic' Yukawa
unification so that the $b$-quark mass can take
acceptable values even with universal boundary
conditions. For $\mu<0$, a suppressed deviation
from Yukawa unification may be adequate, while,
for $\mu>0$, a more sizable deviation is required.
In the $\mu<0$ case, however, the sparticles are
considerably heavier due to the constraints from
$b\rightarrow s\gamma$ and the muon anomalous
magnetic moment. So, the $\mu>0$ case is
more attractive for experimenters.

\par
We considered a particular SUSY GUT from the above
class with a deviation from Yukawa unification
which is adequate for $\mu>0$. We then discussed
the resulting MSSM under the assumption of
universal boundary conditions and the various
phenomenological and cosmological requirements
which restrict its parameter space. They originate
from the data on the inclusive branching ratio of
$b\rightarrow s\gamma$, the muon anomalous magnetic
moment, the CDM abundance in the universe, and the
masses $m_h$ and $m_A$.

\par
The calculation of ${\rm BR}(b\rightarrow s\gamma)$
incorporates all the LO QED and NLO QCD corrections
which hold for large values of $\tan\beta$, while
the LSP contribution to $\Omega_{CDM}~h^2$ is
evaluated by using the code {\tt micrOMEGAs} which
includes all possible coannihilation processes,
treats poles properly and uses the one-loop QCD
corrected Higgs decay widths. We also employed an
alternative method for estimating the LSP relic
abundance and found excellent agreement with
{\tt micrOMEGAs}.

\par
We showed that, in the particular model with Yukawa
quasi-unification considered, there exists a wide
and natural range of CMSSM parameters which is
consistent with all the above constraints. We found
that, within the investigated part of the overall
allowed parameter space, the parameter $\tan\beta$
ranges between about 58 and 61 and the `asymptotic'
splitting between the bottom (or tau) and the top
Yukawa couplings varies in the range $26-35\%$ for
central values of $m_b(M_Z)$ and $\alpha_s(M_Z)$.
Also, the LSP mass can be as low as about
$107~{\rm GeV}$.

\acknowledgements{
We thank T. Falk, G. Ganis, P. Gondolo and
A. Santamaria for useful discussions. We also thank
the {\tt micrOMEGAs} team, G. B\'{e}langer,
F. Boudjema, A. Pukhov and A. Semenov, for their help
in the use of their code and A. Moro for his help in
the graphics. This work was supported by European
Union under the RTN contracts HPRN-CT-2000-00148 and
HPRN-CT-2000-00152. M.E.G. acknowledges support from
the `Funda\c c\~ao para a Ci\^encia e Tecnologia'
under contract SFRH/BPD/5711/2001.}

\def\ijmp#1#2#3{{Int. Jour. Mod. Phys. }{\bf #1~}(#2)~#3}
\def\plb#1#2#3{{Phys. Lett. }{\bf B~#1~}(#2)~#3}
\def\zpc#1#2#3{{Z. Phys. }{\bf C~#1~}(#2)~#3}
\def\prl#1#2#3{{Phys. Rev. Lett. }{\bf #1~}(#2)~#3}
\def\rmp#1#2#3{{Rev. Mod. Phys. }{\bf #1~}(#2)~#3}
\def\prep#1#2#3{{Phys. Rep. }{\bf #1~}(#2)~#3}
\def\prd#1#2#3{{Phys. Rev. }{\bf D~#1~}(#2)~#3}
\def\npb#1#2#3{{Nucl. Phys. }{\bf B~#1~}(#2)~#3}
\def\npps#1#2#3{{Nucl. Phys. (Proc. Sup.) }
{\bf B~#1~}(#2)~#3}
\def\mpl#1#2#3{{Mod. Phys. Lett. }{\bf #1~}(#2)~#3}
\def\arnps#1#2#3{{Annu. Rev. Nucl. Part. Sci. }
{\bf #1~}(#2)~#3}
\def\sjnp#1#2#3{{Sov. J. Nucl. Phys. }{\bf #1~}(#2)~#3}
\def\jetp#1#2#3{{JETP Lett. }{\bf #1~}(#2)~#3}
\def\app#1#2#3{{Acta Phys. Polon. }{\bf #1~}(#2)~#3}
\def\rnc#1#2#3{{Riv. Nuovo Cim. }{\bf #1~}(#2)~#3}
\def\ap#1#2#3{{Ann. Phys. }{\bf #1~}(#2)~#3}
\def\ptp#1#2#3{{Prog. Theor. Phys. }{\bf #1~}(#2)~#3}
\def\apjl#1#2#3{{Astrophys. J. Lett. }{\bf #1~}(#2)~#3}
\def\n#1#2#3{{Nature }{\bf #1~}(#2)~#3}
\def\apj#1#2#3{{Astrophys. Journal }{\bf #1~}(#2)~#3}
\def\anj#1#2#3{{Astron. J. }{\bf #1~}(#2)~#3}
\def\mnras#1#2#3{{MNRAS }{\bf #1~}(#2)~#3}
\def\grg#1#2#3{{Gen. Rel. Grav. }{\bf #1~}(#2)~#3}
\def\s#1#2#3{{Science }{\bf #1~}(#2)~#3}
\def\baas#1#2#3{{Bull. Am. Astron. Soc. }{\bf #1~}(#2)~#3}
\def\ibid#1#2#3{{ibid. }{\bf #1~}(#2)~#3}
\def\cpc#1#2#3{{Comput. Phys. Commun. }{\bf #1~}(#2)~#3}
\def\astp#1#2#3{{Astropart. Phys. }{\bf #1~}(#2)~#3}
\def\epjc#1#2#3{{Eur. Phys. J. }{\bf C~#1~}(#2)~#3}
\def\nima#1#2#3{{Nucl. Instrum. Meth. }{\bf A~#1~}(#2)~#3}
\def\jhep#1#2#3{{JHEP }{\bf #1~}(#2)~#3}

\newpage
\vspace{.1cm}
\pagestyle{empty}
\renewcommand{\arraystretch}{.3}
\baselineskip 12pt

\begin{figure}[tb]
\begin{center}
\epsfig{figure=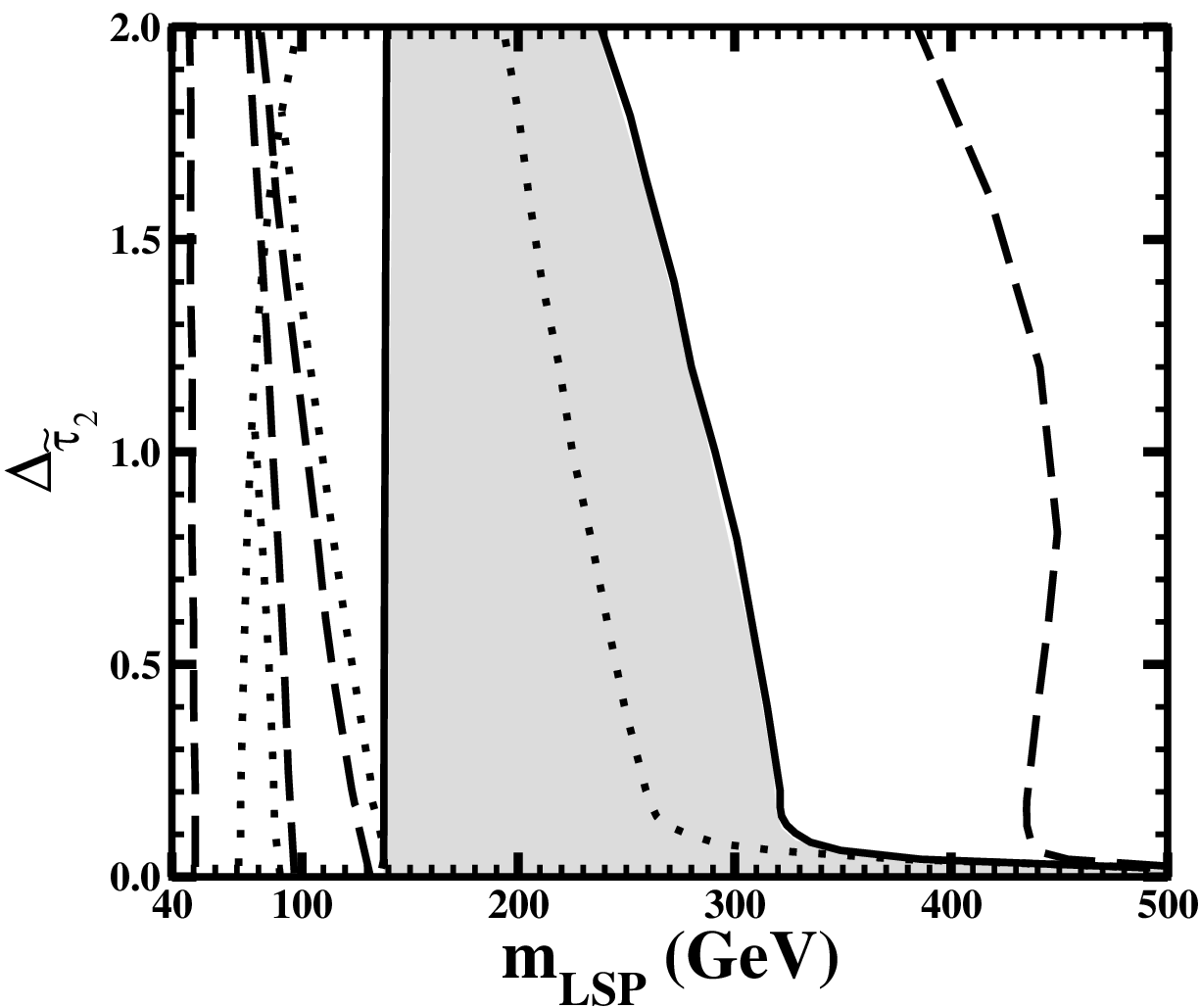,height=3.29in,angle=0}
\end{center}
\medskip
{\footnotesize
\par {FIG. \Figa.} Restrictions on the
$m_{LSP}-\Delta_{\tilde\tau_2}$ plane for $A_0=0$,
$\alpha_s(M_Z)=0.1185$. From left to right, the
dashed (dotted) lines depict the lower bounds on
$m_{LSP}$ from $m_A>110~{\rm GeV}$,
${\rm BR}(b\rightarrow s\gamma)>1.9\times 10^{-4}$
and $\delta a_\mu<58\times 10^{-10}$, and the
upper bound on $m_{LSP}$ from
$\Omega_{LSP}~h^2<0.22$ for
$m_b(M_Z)=2.684~{\rm GeV}$
($3.092~{\rm GeV}$). The left (right) solid
line depicts the lower (upper) bound on $m_{LSP}$
from $m_h>114.1~{\rm GeV}$ ($\Omega_{LSP}~h^2<0.22$)
for $m_b(M_Z)=2.888~{\rm GeV}$. The allowed area for
$m_b(M_Z)=2.888~{\rm GeV}$ is shaded.}
\end{figure}

\begin{figure}[tb]
\begin{center}
\epsfig{figure=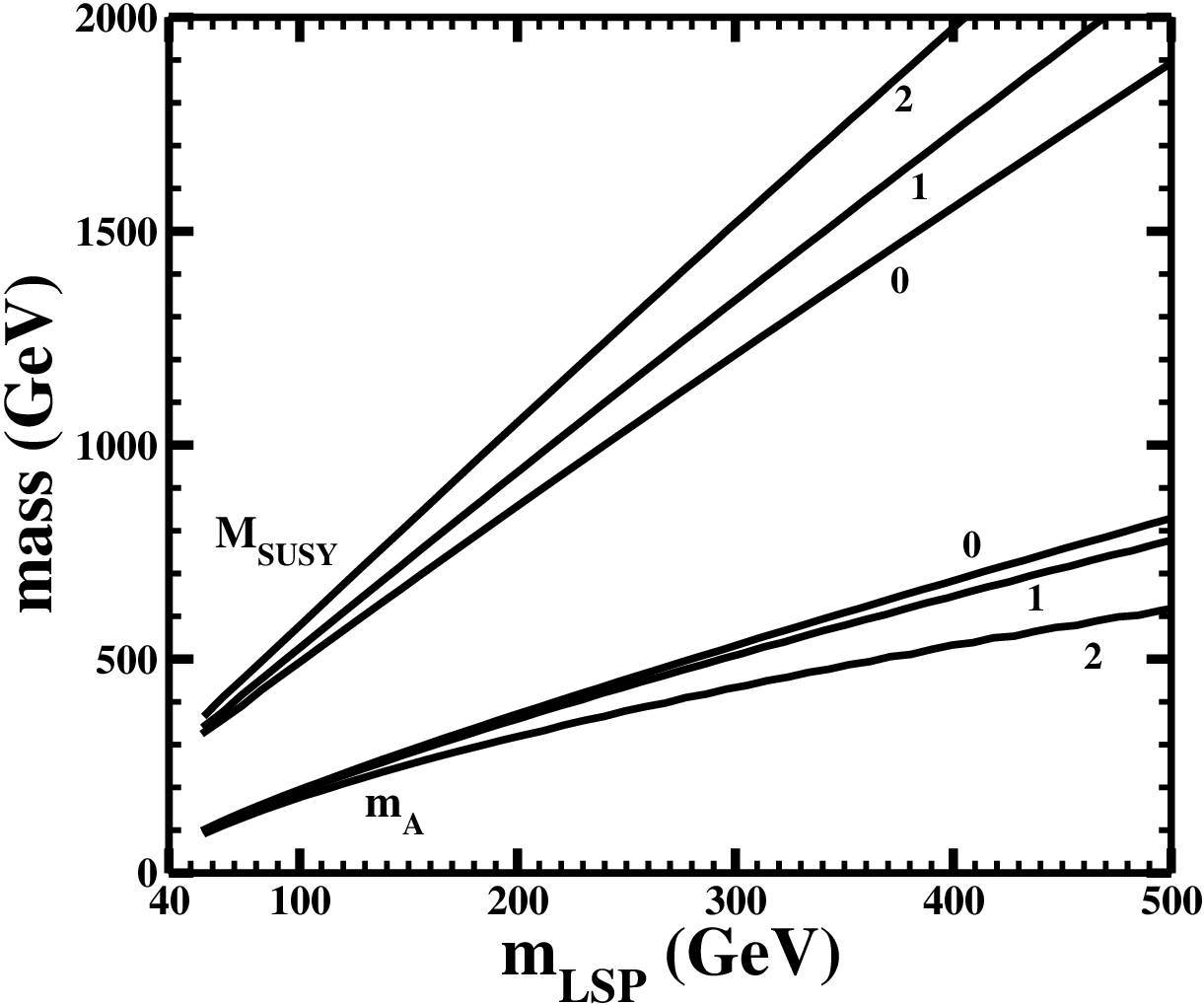,height=3.29in,angle=0}
\end{center}
\medskip
{\footnotesize
\par {FIG. \Figb.} The mass parameters $m_A$ and
$M_{SUSY}$ as functions of $m_{LSP}$ for various
values of $\Delta_{\tilde\tau_2}$, which are
indicated on the curves, and with $A_0=0$,
$m_b(M_Z)=2.888~{\rm GeV}$, $\alpha_s(M_Z)=0.1185$.}
\end{figure}

\begin{figure}[tb]
\begin{center}
\epsfig{figure=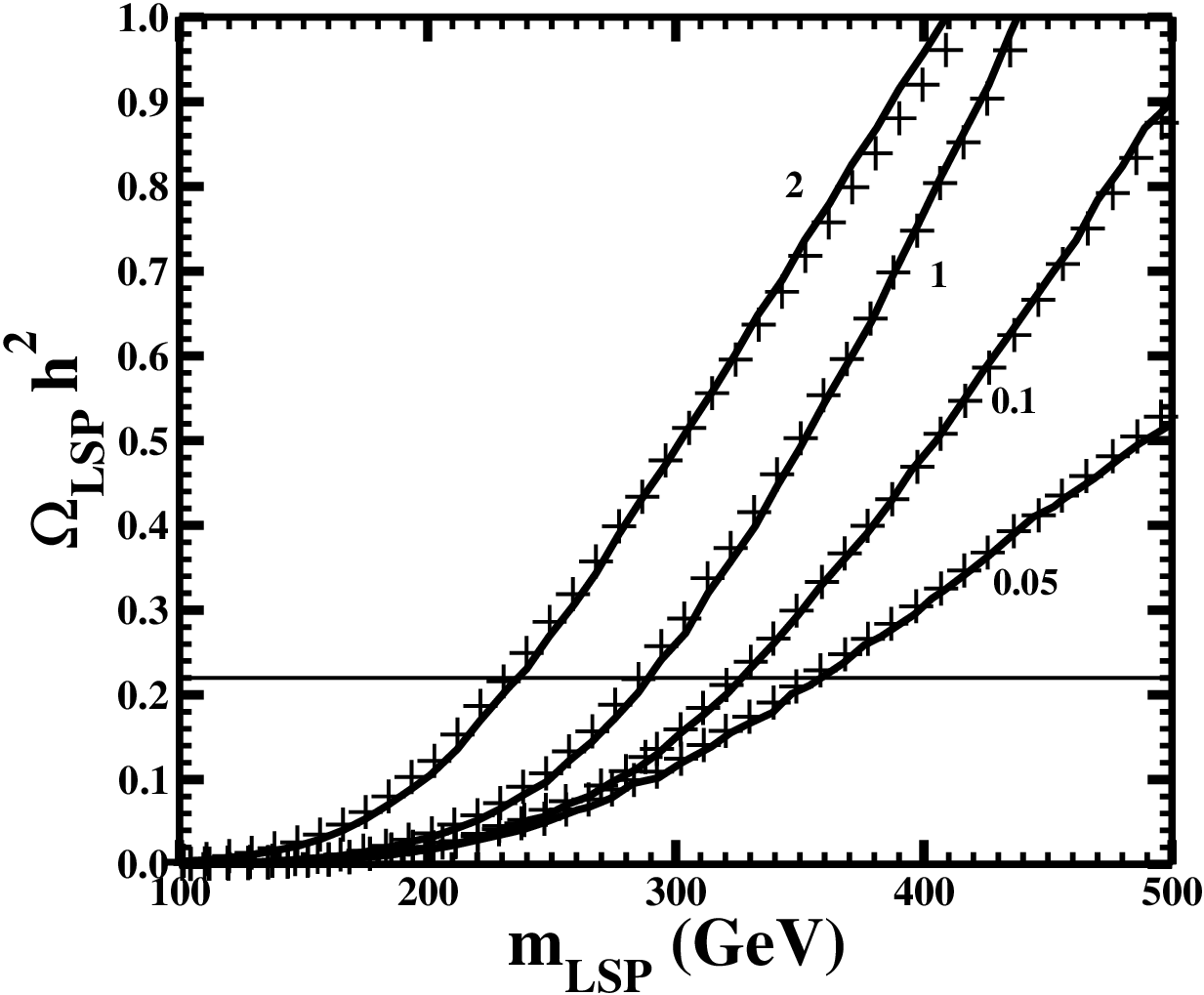,height=3.29in,angle=0}
\end{center}
\medskip
{\footnotesize
\par {FIG. \Figc.} The LSP relic abundance
$\Omega_{LSP}~h^2$ versus $m_{LSP}$ for various
$\Delta_{\tilde\tau_2}$'s (indicated on the curves)
and with $A_0=0$, $m_b(M_Z)=2.888~{\rm GeV}$,
$\alpha_s(M_Z)=0.1185$. The solid lines (crosses)
are obtained by {\tt micrOMEGAs} (our alternative
method). The upper bound on $\Omega_{LSP}~h^2$
(=0.22) is also depicted.}
\end{figure}

\begin{figure}[tb]
\begin{center}
\epsfig{figure=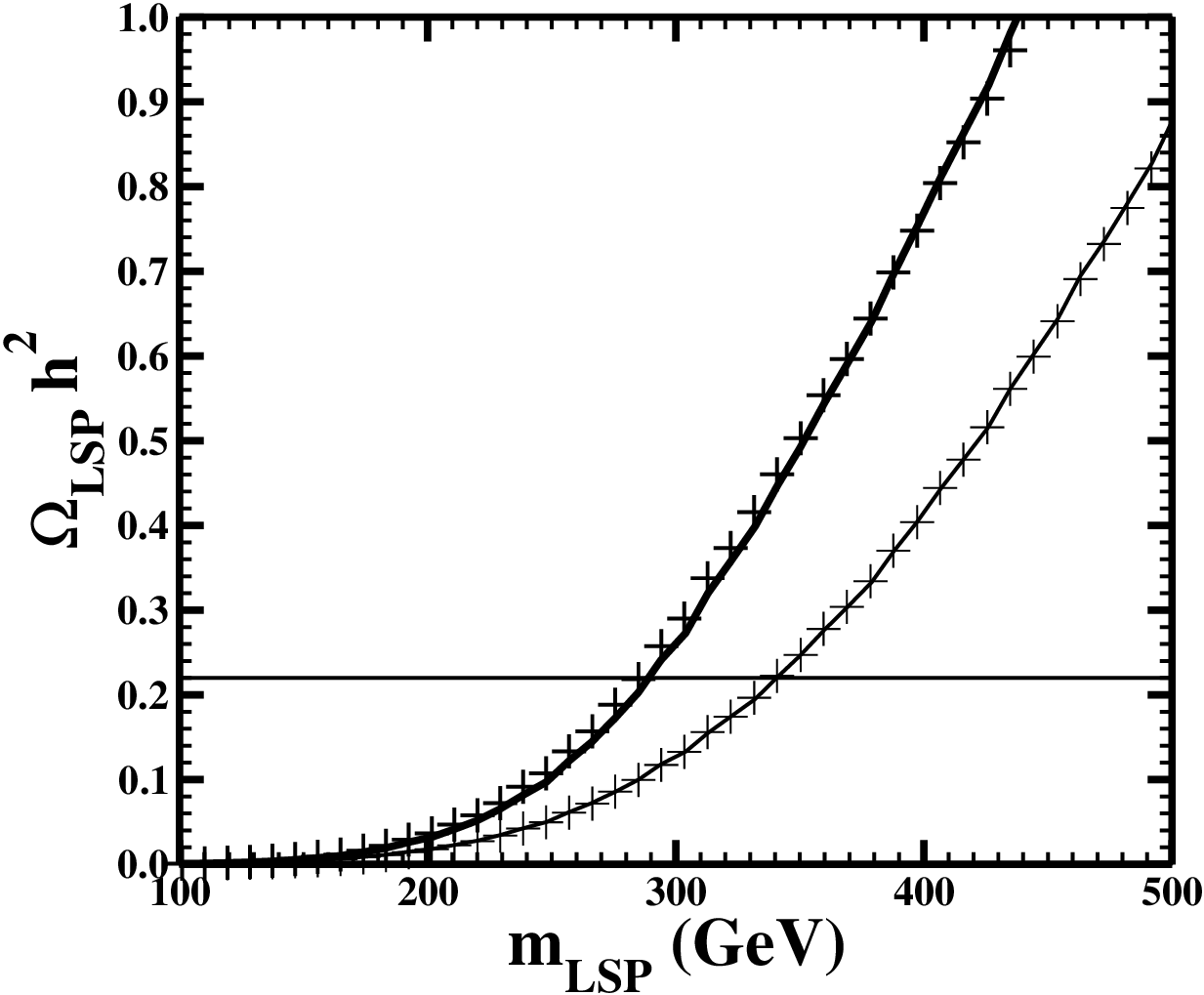,height=3.29in,angle=0}
\end{center}
\medskip
{\footnotesize
\par {FIG. \Figd.} The LSP relic abundance
$\Omega_{LSP}~h^2$ versus $m_{LSP}$ for
$\Delta_{\tilde\tau_2}=1$, $A_0=0$,
$m_b(M_Z)=2.888~{\rm GeV}$, $\alpha_s(M_Z)=0.1185$.
The thick (faint) solid line is obtained by
{\tt micrOMEGAs} with (without) the one-loop QCD
corrections to the Higgs widths, while the thick
(faint) crosses by our alternative method with
(without) the `correction' of the $b$-quark mass.
The upper bound on $\Omega_{LSP}~h^2$ (=0.22) is
also depicted.}
\end{figure}

\begin{figure}[tb]
\begin{center}
\epsfig{figure=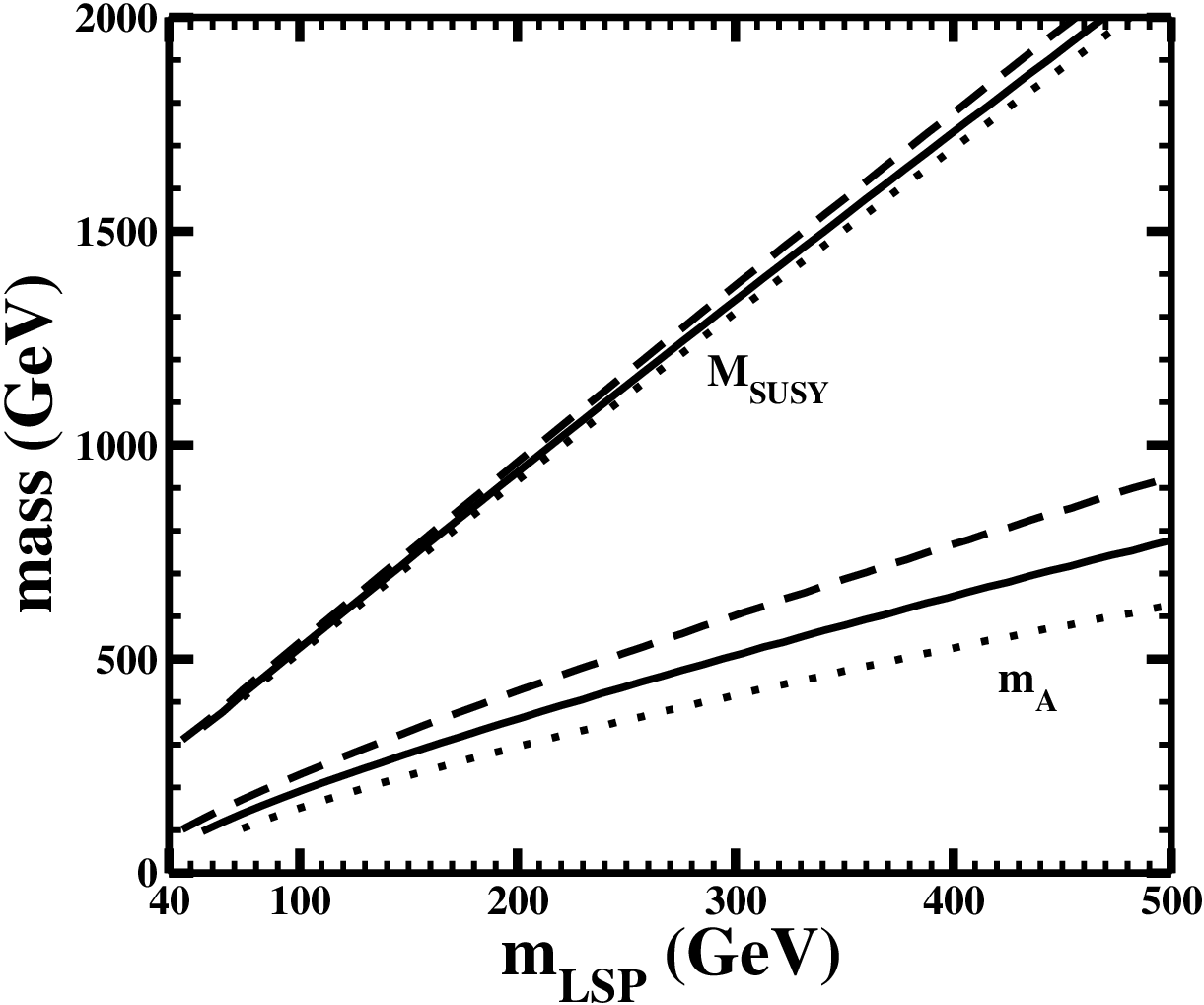,height=3.29in,angle=0}
\end{center}
\medskip
{\footnotesize
\par {FIG. \Fige.} The mass parameters $m_A$ and
$M_{SUSY}$ versus $m_{LSP}$ for
$\Delta_{\tilde\tau_2}=1$, $A_0=0$,
$\alpha_s(M_Z)=0.1185$ and with
$m_b(M_Z)=2.684~{\rm GeV}$ (dashed lines),
$3.092~{\rm GeV}$ (dotted lines) or
$2.888~{\rm GeV}$ (solid lines).}
\end{figure}

\begin{figure}[tb]
\begin{center}
\epsfig{figure=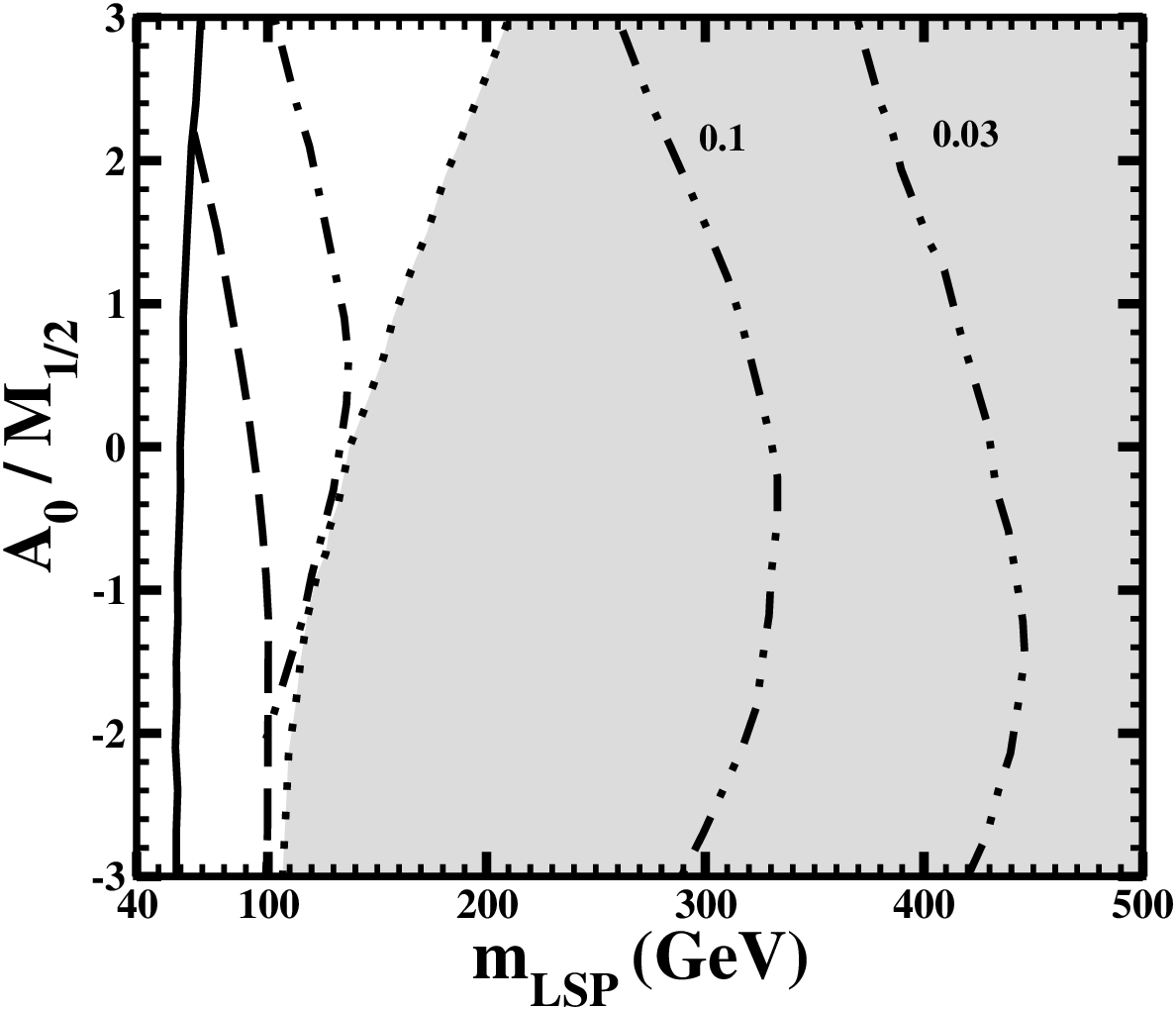,height=3.29in,angle=0}
\end{center}
\medskip
{\footnotesize
\par {FIG. \Figf.} Restrictions on the
$m_{LSP}-A_0/M_{1/2}$ plane for
$\Delta_{\tilde\tau_2}=0$, $m_b(M_Z)=2.888~{\rm GeV}$,
$\alpha_s(M_Z)=0.1185$. The solid, dashed,
dot-dashed and dotted lines correspond to the lower
bounds on $m_{LSP}$ from $m_A>110~{\rm GeV}$,
${\rm BR}(b\rightarrow s\gamma)>1.9\times 10^{-4}$,
$\delta a_\mu<58\times 10^{-10}$ and
$m_h>114.1~{\rm GeV}$ respectively. The upper bound
on $m_{LSP}$ from $\Omega_{LSP}~h^2<0.22$ does not
appear in the figure since it lies at
$m_{LSP}>500~{\rm GeV}$. The allowed area is shaded.
For comparison, we also display the bounds from
$\Omega_{LSP}~h^2<0.22$ (double dot-dashed lines)
for $\Delta_{\tilde\tau_2}=0.1~{\rm and}~0.03$, as
indicated.}
\end{figure}

\begin{figure}[tb]
\begin{center}
\epsfig{figure=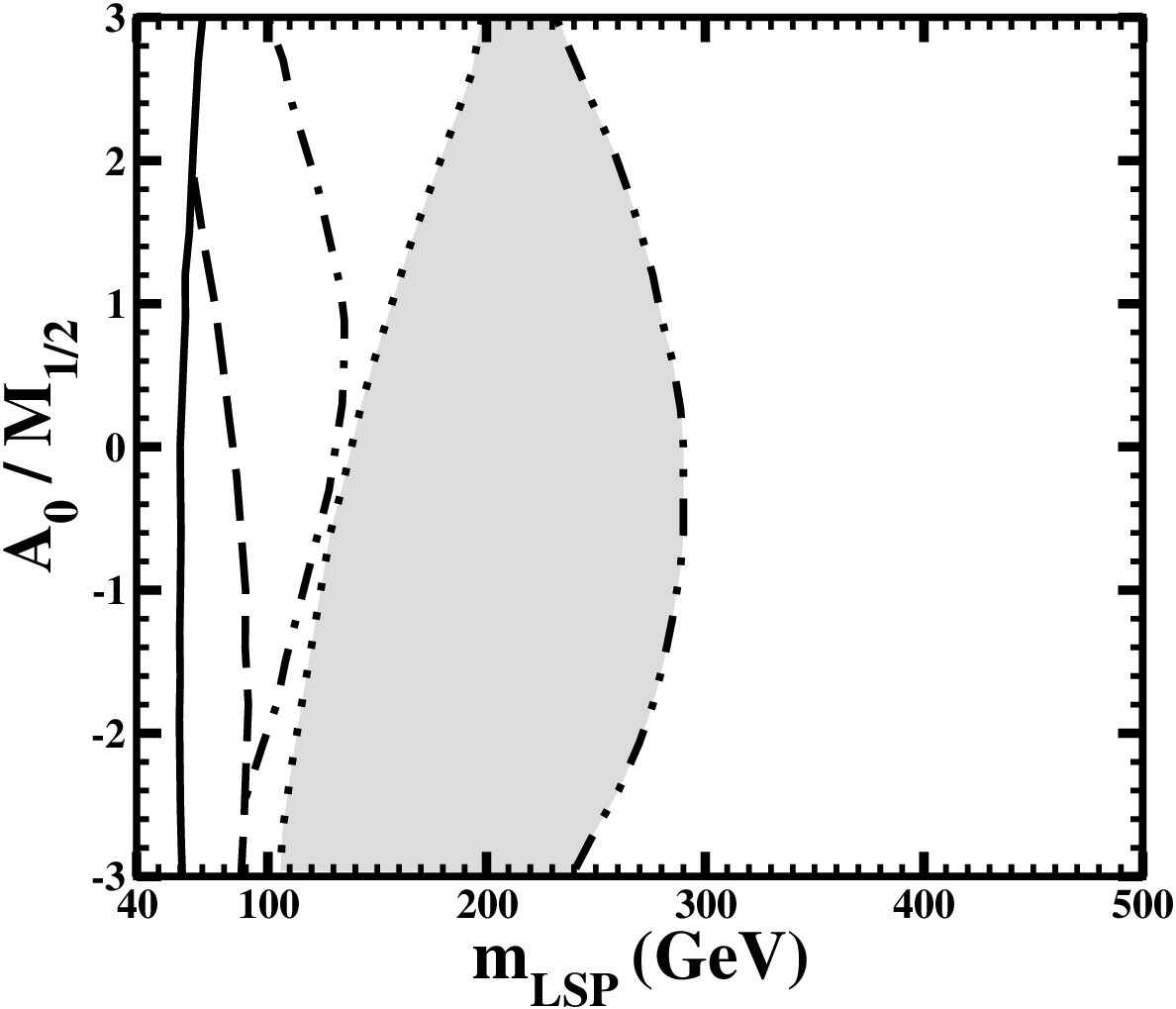,height=3.29in,angle=0}
\end{center}
\medskip
{\footnotesize
\par {FIG. \Figg.} Restrictions on the
$m_{LSP}-A_0/M_{1/2}$ plane for
$\Delta_{\tilde\tau_2}=1$, $m_b(M_Z)=2.888~{\rm GeV}$
and $\alpha_s(M_Z)=0.1185$. We use the same notation
for the lines which correspond to the various bounds
on $m_{LSP}$ and for the allowed area as in Fig.\Figf.
The upper bound on $m_{LSP}$ from the cosmological
constraint $\Omega_{LSP}~h^2<0.22$ (double dot-dashed
line) now appears inside the figure.}
\end{figure}

\begin{figure}[tb]
\begin{center}
\epsfig{figure=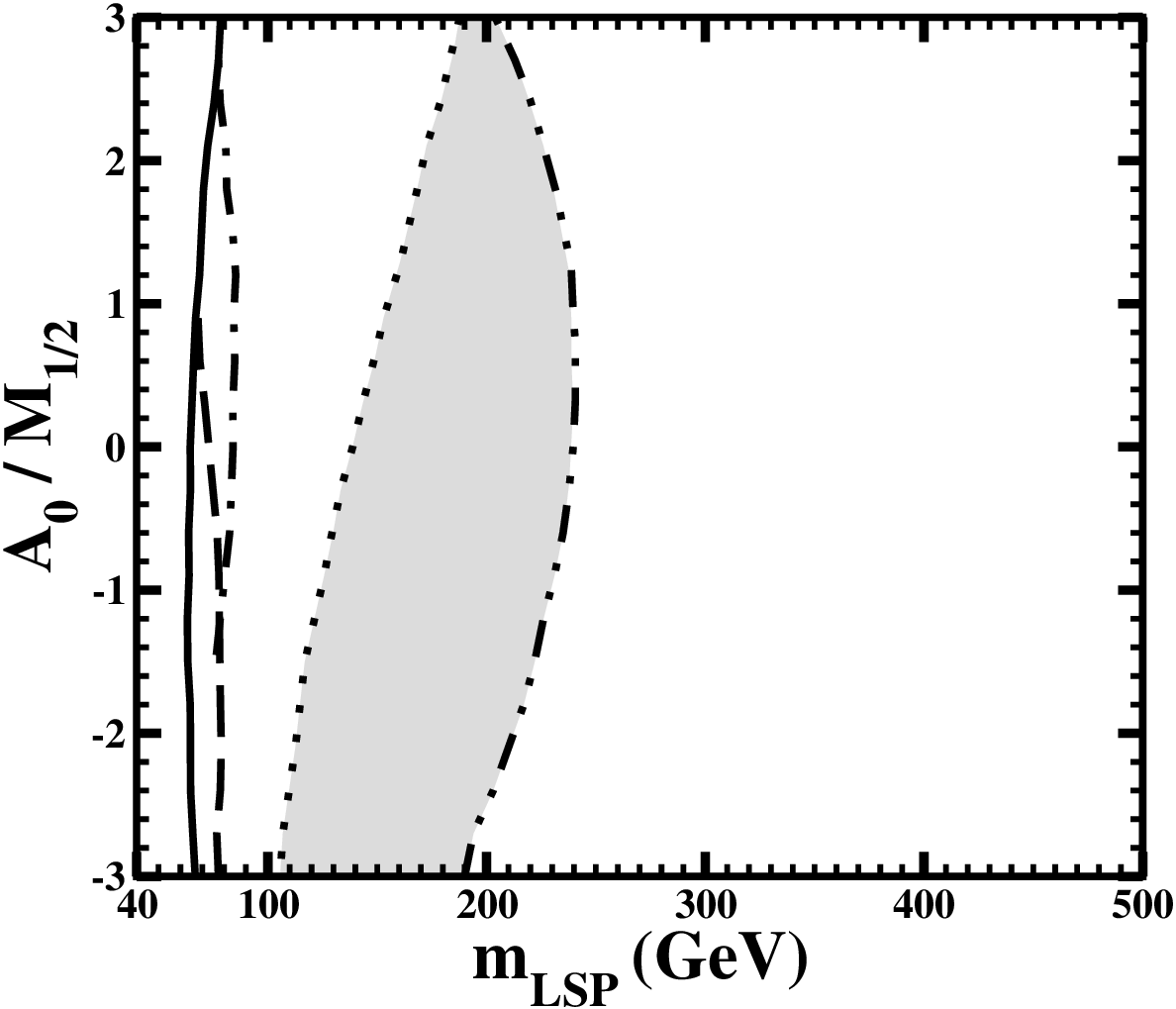,height=3.29in,angle=0}
\end{center}
\medskip
{\footnotesize
\par {FIG. \Figh.} Restrictions on the
$m_{LSP}-A_0/M_{1/2}$ plane for
$\Delta_{\tilde\tau_2}=2$, $m_b(M_Z)=2.888~{\rm GeV}$
and $\alpha_s(M_Z)=0.1185$. We use the same notation
for the lines which correspond to the various bounds
on $m_{LSP}$ and for the allowed area as in Fig.\Figg.}
\end{figure}

\begin{figure}[tb]
\begin{center}
\epsfig{figure=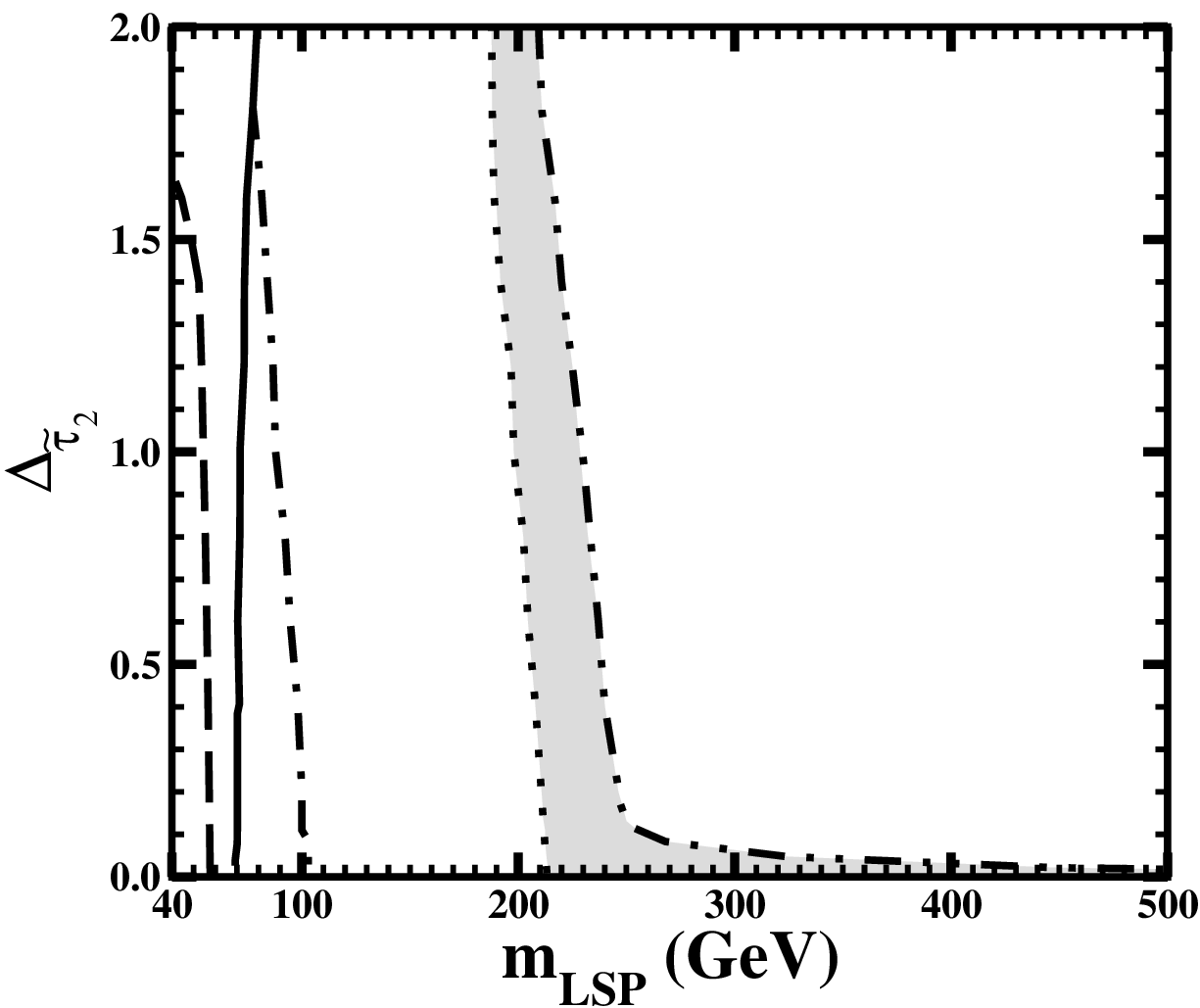,height=3.29in,angle=0}
\end{center}
\medskip
{\footnotesize
\par {FIG. \Figi.} Restrictions on the
$m_{LSP}-\Delta_{\tilde\tau_2}$ plane for
$A_0/M_{1/2}=3$, $m_b(M_Z)=2.888~{\rm GeV}$ and
$\alpha_s(M_Z)=0.1185$. We use the same notation
for the lines which correspond to the various
bounds on $m_{LSP}$ and for the allowed area as
in Fig.\Figg.}
\end{figure}

\begin{figure}[tb]
\begin{center}
\epsfig{figure=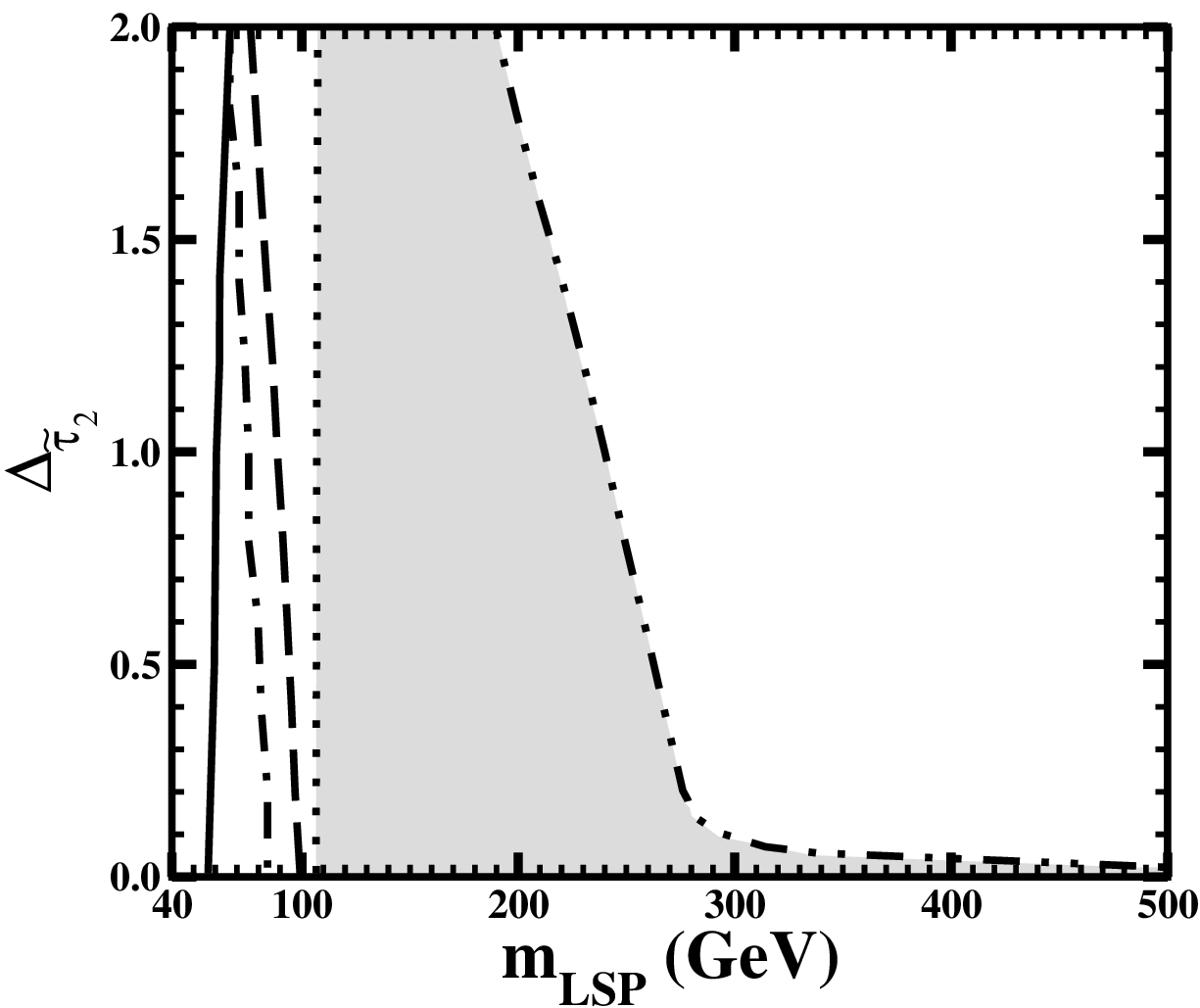,height=3.29in,angle=0}
\end{center}
\medskip
{\footnotesize
\par {FIG. \Figj.} Restrictions on the
$m_{LSP}-\Delta_{\tilde\tau_2}$ plane for
$A_0/M_{1/2}=-3$, $m_b(M_Z)=2.888~{\rm GeV}$ and
$\alpha_s(M_Z)=0.1185$. We use the same notation
for the lines which correspond to the various
bounds on $m_{LSP}$ and for the allowed area as
in Fig.\Figg. The minimal value of $m_{LSP}$ in
the investigated overall allowed parameter space
is about $107~{\rm GeV}$ for central values of
$m_b(M_Z)$ and $\alpha_s(M_Z)$. As seen from this
figure, this value is achieved at $A_0/M_{1/2}=-3$
and practically any $\Delta_{\tilde\tau_2}$
(see also Figs.\Figf, \Figg\ and \Figh).}
\end{figure}


\begin{references}

\bibitem{als} B. Ananthanarayan, G. Lazarides and
Q. Shafi, \prd{44}{1991}{1613}; \plb{300}{1993}{245}.
For a more recent update see U. Sarid, hep-ph/9610341.

\bibitem{pana} G. Lazarides and C. Panagiotakopoulos,
\plb{337}{1994}{90}.

\bibitem{copw} M. Carena, M. Olechowski,
S. Pokorski and C.E.M. Wagner, Nucl. Phys. {\bf B~426}
(1994) 269; M. Carena and C.E.M. Wagner, hep-ph/9407209.

\bibitem{hall} L. Hall, R. Rattazzi and U. Sarid,
\prd{50}{1994}{7048};
R. Hempfling, \prd{49}{1994}{6168}.

\bibitem{cd2} M.E. G\'{o}mez, G. Lazarides and C. Pallis,
\plb{487}{2000}{313}.

\bibitem{mb} C.T. Sachrajda, \nima{462}{2001}{23}.

\bibitem{cmssm} J. Ellis, T. Falk, G. Ganis, K. Olive
and M. Srednicki, \plb{510}{2001}{236}.

\bibitem{santamaria} G. Rodrigo and A. Santamaria,
\plb{313}{1993}{441}; G. Rodrigo, A. Pich and
A. Santamaria, \plb{424}{1998}{367}.

\bibitem{rge} F.A. Chishtie and V. Elias,
\prd{64}{2001}{016007}.

\bibitem{dr} B.C. Allanach and S.F. King,
\npb{507}{1997}{91}.

\bibitem{wetterich} G. Lazarides, Q. Shafi and
C. Wetterich, \npb{181}{1981}{287}; G. Lazarides and
Q. Shafi, \npb{350}{1991}{179}.

\bibitem{cleo} R. Barate {\it et al.} (ALEPH
Collaboration), \plb{429}{1998}{169}; K. Abe
{\it et al.} (BELLE Collaboration),
\plb{511}{2001}{151}; S. Chen {\it et al.} (CLEO
Collaboration), \prl{87}{2001}{251807}.

\bibitem{muon} H.N. Brown {\it et al.},
\prl{86}{2001}{2227}.

\bibitem{nonuniversal} H. Baer, M. Brhlik,
M.A. D\'{\i}az, J. Ferrandis, P. Mercadante,
P. Quintana and X. Tata, \prd{63}{2001}{015007};
S.F. King and M. Oliveira, \prd{63}{2001}{015010};
H. Baer and J. Ferrandis, \prl{87}{2001}{211803};
T. Bla\v{z}ek, R. Derm\'{\i}\v{s}ek and S. Raby,
\prl{88}{2002}{111804}; \prd{65}{2002}{115004};
U. Chattopadhyay, A. Corsetti and P. Nath,
hep-ph/0201001.

\bibitem{jean} R. Jeannerot, S. Khalil, G. Lazarides
and Q. Shafi, \jhep{0010}{2000}{012}.

\bibitem{talks} G. Lazarides, in {\it Recent
Developments in Particle Physics and Cosmology},
edited by G.C. Branco, Q. Shafi and J.I. Silva-Marcos
(Kluwer Acad. Pub., Dordrecht, 2001) p. 399
(hep-ph/0011130);
R. Jeannerot, S. Khalil and G. Lazarides, in {\it The
Proceedings of Cairo International Conference on High
Energy Physics}, edited by S. Khalil, Q. Shafi and
H. Tallat (Rinton Press, Princeton, 2001) p. 254
(hep-ph/0106035).

\bibitem{leontaris} I. Antoniadis and G.K. Leontaris,
\plb{216}{1989}{333}.

\bibitem{rsym} G. Lazarides and Q. Shafi,
\prd{58}{1998}{071702}.

\bibitem{cdm} M.E. G\'{o}mez, G. Lazarides and
C. Pallis, \prd{61}{2000}{123512}.

\bibitem{pierce} D. Pierce, J. Bagger, K. Matchev and
R. Zhang, \npb{491}{1997}{3}.

\bibitem{fh} S. Heinemeyer, W. Hollik and G. Weiglein,
hep-ph/0002213.

\bibitem{drees} M. Drees and M.M. Nojiri,
\prd{45}{1992}{2482}.

\bibitem{kagan} A.L. Kagan and M. Neubert,
\epjc{7}{1999}{5}.

\bibitem{nlohiggs} M. Ciuchini, G. Degrassi,
P. Gambino and G. Giudice, \npb{527}{1998}{21}.

\bibitem{nlosusy} G. Degrassi, P. Gambino and
G.F. Giudice, \jhep{0012}{2000}{009}.

\bibitem{carena} M. Carena, D. Garcia, U. Nierste
and C.E.M. Wagner, \plb{499}{2001}{141}.

\bibitem{gambino} P. Gambino and M. Misiak,
\npb{611}{2001}{338}.

\bibitem{gerardo} G. Ganis, private communication.

\bibitem{ganis} J. Ellis, T. Falk, G. Ganis and
K.A. Olive, \prd{62}{2000}{075010}.

\bibitem{marciano} A. Czarnecki and W.J. Marciano,
\prd{64}{2001}{013014}.

\bibitem{davier} M. Davier and A. H\"{o}cker,
\plb{435}{1998}{427}.

\bibitem{sign} M. Knecht and A. Nyffeler,
\prd{65}{2002}{073034}; M. Knecht, A. Nyffeler,
M. Perrottet and E. de Rafael,
\prl{88}{2002}{071802};
M. Hayakawa and T. Kinoshita, hep-ph/0112102;
I. Blokland, A. Czarnecki and K. Melnikov,
\prl{88}{2002}{071803}.

\bibitem{gminus2} T. Ibrahim and P. Nath,
\prd{61}{2000}{015004}; S.P. Martin and J.D. Wells,
\prd{64}{2001}{035003}.

\bibitem{dasi} C. Pryke {\it et al.},
\apj{568}{2002}{46}.

\bibitem{micro} G. B\'{e}langer, F. Boudjema,
A. Pukhov and A. Semenov, hep-ph/0112278.

\bibitem{baer} H. Baer, C. Bal\'{a}zs and
A. Belyaev, \jhep{0203}{2002}{042}.

\bibitem{coan} K. Griest and D. Seckel,
\prd{43}{1991}{3191}.

\bibitem{width} A. Djouadi, M. Spira and P.M. Zerwas,
\zpc{70}{1996}{427}; A. Djouadi, J. Kalinowski and
M. Spira, \cpc{108}{1998}{56}.

\bibitem{susy} C. Bal\'{a}zs, J.L. Diaz-Cruz,
H.-J. He, T. Tait and C.-P. Yuan, Phys. Rev.
{\bf D~59} (1999) 055016; M. Carena, S. Mrenna and
C.E.M. Wagner, \prd{60}{1999}{075010}; M. Carena,
D. Garcia, U. Nierste and C.E.M. Wagner,
\npb{577}{2000}{88}.

\bibitem{darksusy} P. Gondolo, J. Edsj\"{o},
L. Bergstr\"{o}m, P. Ullio and E.A. Baltz,
astro-ph/0012234.

\bibitem{roberto} T. Nihei, L. Roszkowski and
R. Ruiz de Austri, \jhep{0203}{2002}{031}.

\bibitem{ellis} J. Ellis, T. Falk and K.A. Olive,
\plb{444}{1998}{367}; J. Ellis, T. Falk, K.A. Olive
and M. Srednicki, \astp{13}{2000}{181},
(E) \ibid{15}{2001}{413}.

\bibitem{hw} S. Khalil, G. Lazarides and C. Pallis,
\plb{508}{2001}{327}.

\bibitem{arnowitt} R. Arnowitt, B. Dutta and
Y. Santoso, \npb{606}{2001}{59}.

\bibitem{vergados} M.E. G\'{o}mez and J.D. Vergados,
\plb{512}{2001}{252}.

\bibitem{falk} T. Falk, private communication.

\bibitem{higgs} ALEPH, DELPHI, L3 and OPAL
Collaborations, The LEP working group for Higgs
boson searches, hep-ex/0107029.

\bibitem{cdf} J.A. Valls (CDF Collaboration),
FERMILAB-CONF-99/263-E.

\end{references}
\end{document}